\documentclass[prd,referee,showpacs,nofootinbib,twocolumn]{revtex4-1}
\usepackage{amsmath}
\usepackage{eurosym}
\usepackage{amssymb}
\usepackage{graphicx}
\usepackage{color}

\newcommand{\bea}{\begin{eqnarray}}
\newcommand{\eea}{\end{eqnarray}}
\newcommand{\be}{\begin{equation}}
\newcommand{\ee}{\end{equation}}

\begin{document}

\title{Loop Quantum Cosmology with a noncommutative quantum deformed
photon gas}
\author{Yunxin Ye}
\email{yeyunxin96@hotmail.com}
\affiliation{School of Physics, Sun Yat-Sen University, Guangzhou 510275, People's
Republic of China}
\author{Tiberiu Harko}
\email{t.harko@ucl.ac.uk}
\affiliation{Department of Physics, Babes-Bolyai University, Kogalniceanu Street,
Cluj-Napoca 400084, Romania}
\affiliation{School of Physics, Sun Yat-Sen University,   Guangzhou 510275, People's
Republic of China}
\affiliation{Department of Mathematics, University College London, Gower Street, London
WC1E 6BT, United Kingdom}
\author{Shi-Dong Liang}
\email{stslsd@mail.sysu.edu.cn}
\affiliation{School of Physics, Sun Yat-Sen University,   Guangzhou 510275, People's
Republic of China,}
\affiliation{State Key Laboratory of Optoelectronic Material and Technology, and
Guangdong Province Key Laboratory of Display Material and Technology,
Sun Yat-Sen University,   Guangzhou 510275, People's Republic of China}

\begin{abstract}
 The noncommutativity of the space-time had important implications for the very early Universe, when its size was of the order of the Planck length. An important implication of this effect is the deformation of the standard dispersion relation of special
relativity. Moreover, in the Planck regime gravity itself must be described by a quantum theory. We consider the implications of the modified dispersion relations for a photon gas, filling the early Universe, in the framework of Loop Quantum Cosmology, a theoretical approach to quantum gravity. We consider three types of deformations of the dispersion relations of the photon gas,  from which we obtain the  Planck scale corrections to the energy density and pressure. The cosmological implications of the modified equations of state are explored in detail for all radiation models in the framework of the modified Friedmann equation of Loop Quantum Cosmology.  By numerically integrating the evolution equations we investigate the evolution of the basic cosmological parameters (scale factor, Hubble function, radiation temperature, and deceleration parameter) for a deformed photon gas filled Universe. In all models the evolution of the Universe shows the presence of a (nonsingular) bounce, corresponding to the transition from a contracting to an expanding phase.
\end{abstract}

\maketitle
\tableofcontents

\section{Introduction}

Cosmology is undergoing presently a very rapid development.  Very
precise observational data, obtained from the study of both Cosmic Microwave
Background (CMB) \cite{c1,c2,c3}, and the matter distribution in the Universe \cite{c4,c5}, did offer the
possibility of testing sophisticated cosmological models. Moreover, the high
level of accuracy of cosmological data allows the testing of some of the
predictions of elementary particle physics, and string theory.

The inflationary cosmological paradigm, introduced  in 1981 by Alan Guth \cite{Guth} is one of the cornerstones of modern cosmology. It attempts to give a solution to some of the major, and yet unsolved, problems of Standard Big Bang Cosmology (horizon, flatness, magnetic monopole, entropy etc.)  by postulating the existence, during the very early period of the existence of the Universe, of an epoch of de Sitter type exponential expansion. Inflation also provides a theoretical explanation for the origin of inhomogeneities in the Universe \cite{c6,c7,c8,c9,c10}, and it predicts that the spectrum of the primordial perturbations
is very close to scale invariance, a result which is consistent with present day observations \cite{c3}. Hence the primordial power spectrum predicted by inflation can thus explain the late formation of large scale structures in the Universe, as well as the small inhomogeneities present in the Cosmic Microwave Background.  For reviews of the inflationary paradigm and of the corresponding cosmological models of the early Universe see \cite{rev1,rev2,rev3,rev4,rev5}.

However, the inflationary paradigm also faces  a number of fundamental theoretical problems, like, for example, the initial singularity problem, and the trans-Planckian problem, which originated from the fact  that the origin and early
evolution of the cosmological fluctuations must have taken place in the trans-Planckian regime, where both General Relativity as well as quantum field theory break down \cite{B1}. For a review of the fundamental conceptual problems that confront inflationary cosmology see \cite{UC0, UC}.

In the standard cosmological models based on the theory of general relativity a fundamental assumption is that (ordinary) matter satisfies the standard energy conditions. In this case the Penrose-Hawking singularity theorem \cite{P1,P2} is valid, and therefore every expanding cosmological solution (isotropic or anisotropic, homogeneous or inhomogeneous) of general relativity must had began with a singularity. However, from a physical point of view, the existence of the strong curvature singularity may be just the result of the extension of general relativity, an essentially classical theory,  beyond its domain of applicability. According to the basic rules of quantum theory, general relativity, as a classical theory, must break down at length and time scales smaller than the Planck length and time, respectively. Hence, assuming the correctness of the quantum theory, it follows that general relativity cannot correctly describe the physics near the singular state.

To solve some of these theoretical problems raised by the description of the early Universe, an alternative scenario, called Bouncing Cosmology, was developed within the framework of the classical theory of gravity \cite{BC1,SolutionHPFP, BC2, BC3} (for recent reviews see \cite{B1, ReviewOnBouncing1a, ReviewOnBouncing1b, ReviewOnBouncing1c,  ReviewOnBouncing2}).  There are two basic approaches to Bouncing Cosmology, the introduction of a nontrivial stress-energy tensor in the matter sector of the Einstein field equations, or the modification of the gravitational Lagrangian in the Einstein-Hilbert action. The classical bouncing scenarios  solve some of the puzzles of the standard model of cosmology \cite{SolutionHPFP}, and they also  can alleviate the singularity problem by postulating that the Universe begins its evolution with a contracting phase, and then it bounces back to the expanding phase before attaining the Planck length. However, these model also have serious drawbacks, like, for example, the instability problem \cite{ ReviewOnBouncing2}. But, most importantly, since the Universe attains extremely high densities very early, it follows that quantum gravity effects would necessarily become extremely important at such small length scales, and high energies. Therefore, a quantum gravitational description  must be necessarily used to explore the high density regimes. Hence, if one would to solve the singularity problem, the use of quantum gravity and/or quantum mechanics becomes inevitable.

One of the promising candidates for a quantum gravity theory is Loop Quantum Gravity (LQG),  whose starting point is the Ashtekar formalism of general relativity and the introduction of loop presentation \cite{lqg1}. LQG starts from the fundamental assumption that even at the quantum level the gravitational force  is an expression of the geometry of the spacetime. From this postulate  LQG advances further and develops a systematic and mathematically rigorous theory of quantum geometry, with a specific emphasis on the Riemannian one. For an in depth introduction to LQG see \cite{Thiemann1}. Generally, LQG represents a nonperturbative background approach to quantize gravity \cite{QuantumGravity1a, QuantumGravity1b,QuantumGravity1c}. The theory fully takes into account the possible existence of fundamental discrete structures at the Planck scale, where all the geometric observables, including physical surfaces and volumes of bounded regions, are essentially quantum in their nature \cite{Asthekar1, Asthekar2, Bianchi1}.

The application of the LQG quantization methods to  spacetimes with reduced symmetry, and in particular to homogenous and isotropic spacetimes, led to the development of the field of Loop Quantum Cosmology (LQC) theory \cite{Bojowald1}. One of the major successes of LQC is the possibility of solving the singularity problem. This is due to the important result according to which the macroscopic physical observables, like, for example,  the energy density, and the curvature, which in general relativity do diverge at the initial moment of the Big Bang,  have a finite upper bound in LQC. The finite nature of the macroscopic quantities follows from the fact that they depend on  the fundamental microscopic parameter of the LQC theory, the fundamental area gap, whose smallest eigen value is always nonzero \cite{Asthekar3, Asthekar4, Asthekar5}. Consequently,  since a maximum value of the energy density can be obtained in LQC, a contracting Friedmann-Lemaitre-Robertson-Walker) (FLRW) Universe will bounce back to an expanding Universe, thus preventing the presence of a physical singularity. It is important to note that this result is derived without introducing any nontrivial or exotic forms of matter into the cosmological models. This represents an important difference as compared to the case of the classical bouncing scenarios. The quantum bounce is generated by the quantum geometrical effects that act as a novel repulsive force, a result that follows immediately from the quantum corrected Friedmann and Raychaudhuri equations. Also, it is important to point out that in the framework of LQC in all classes of spacetimes with different sets of symmetries, including Bianchi and Gowdy models, the singularity problem is successfully solved in most cases \cite{nref1}-\cite{nref4}.  The resolution of sudden singularities, corresponding to the case when the pressure diverges, but the energy density approaches a finite value depends on the ratio of the latter to a critical energy density of the order of the Planck density \cite{nref1}. On the other hand if the value of this ratio is greater than unity, the Universe avoids the sudden future singularity, and enters into an oscillatory phase.

On the other hand one can extend the Loop Quantum Cosmology matter bounce scenario by including a dark energy era, which ends abruptly at a Big Rip singularity, where both the scale factor and the Hubble function do diverge \cite{nref5}. The scalar tensor reconstruction method in LQC as well as some scenarios for which the Hubble function becomes unbounded at finite time, were discussed in \cite{nref6}. This corresponds to a case in which a Big Rip occurs. For reviews of the solution of the singularity problem in LQC see \cite{Asthekar9} and \cite{Singh3}.

In addition to the singularity problem, cosmological models face an other important challenge. In order for the inflationary Universe models to be consistent with the present day observations, the Universe must have expanded at least 50 e-folds. On the other hand, if, for example,  the Universe had expanded more than 70 e-folds, (a value that is required by a large class of inflationary cosmological models \cite{5}), it follows that the wavelengths of all fluctuation modes, which currently are inside the Hubble radius, must have been smaller than the Planck length  at the beginning  of the inflationary era. This represents the so-called trans-Planckian problem \cite{6}.  Hence, the assumption that spacetime can still be treated classically, but that matter fields are essentially quantum in nature cannot be maintained.  Therefore  a quantum treatment of geometry must be also necessary.  Moreover, the dynamics of the Universe before the beginning of the inflation cannot be neglected anymore, even when the modes were well inside the Hubble radius \cite{6,8,Zhu1,Zhu2}.

Hence all these problems requires to look for a quantum theory of gravity and cosmology.  In this framework  LQC deals successfully with one of the major problems in cosmology, by replacing the singularity with a quantum bounce, followed by a  slow roll inflation \cite{18}. Presently, it is already possible to look for experimental tests of LQC,  and to search for observational signatures of the quantum bounce and of the pre-inflationary dynamics in current or forthcoming observations \cite{obs1,obs2,obs3}. Calculations of cosmological perturbations for dressed metrics and for deformed algebras
 \cite{Asthekar9,14} have been performed and investigated numerically in \cite{10,11}. As a result it was found, for example,  that the approach based on deformed algebras is inconsistent with some of the present day observations \cite{21}.

The assumption of the quantum nature of the space-time at the Big Bang immediately raises the problem of its non-commutativity. Theoretical  models derived within the framework of space-time non-commutativity are of particular
theoretical importance.  In 1947, space-time non-commutativity was already suggested by Snyder \cite{snyder}. Snyder put forward a model to quantize space-time, by defining space and time coordinate as operators that keep invariant a five-dimensional quadratic form when operating on it. The five-dimensional quadratic form considered by Snyder can be regarded as a real four-dimensional space with constant curvature, which is a De Sitter space \cite{Desitter}. The theory suggested by Snyder is invariant under Lorentz transformations, however, it is not invariant under space-time translations. Later, C. N. Yang \cite{yang} proposed another theory of quantization of space-time, by defining $L_{i}/\hbar $, in which $L_{i}$ is the angular momentum, $M_{i}/\hbar $, in which $M_{i}$ is the invariance quantity corresponding to Lorentz boost, $Rp_{\mu}/\hbar $, in which $p_{\mu}$ is the momentum 4-vector and $x_{\mu}/a $, in which $x_{\mu}$ is the coordinate 4-vector, as the 15 operators that do not change a six-dimensional quadratic form when operating on it, thus maintaining both Lorentz invariance and translation invariance when implementing discreteness to space-time. Recently, it has been found that space-time non-commutativity is a very promising way of solving some of
the fundamental problems of current physics \cite{Con}. It plays an important role in the description of space-time at scales comparable to the Planck length \cite{Con}. Space-time noncommutativity determines the existence of a minimum length scale, with important implications  in quantum-mechanics, quantum-electrodynamics, thermodynamics, black-hole physics and cosmology \cite{mls, qm1, qm2, grb} etc. Its connection with varying speed of light theories was investigated  in \cite{vsl}.  Space-time noncommutativity also leads to modified energy-momentum dispersion relations, with important implications in physics, cosmology, and astrophysics \cite{ue, pg1, pg2, bhr, wd}. For example, the thermodynamics of the radiation gas with modified dispersion relation $E^{2}-k^{2}=m^{2}\left(1-E/\kappa\right)^{2}$, where $E$ is the photon energy, $k$ is its momentum and $\kappa$ is a constant parameter was studied in \cite{pg1}, while the implications on the thermodynamics of the modified dispersion relation $E^{2}=k^{2}\left[1-\alpha \left(E/E_{P}\right)^{n}\right]$, in which $\alpha $ and $E_{p}$ are constant parameters were studied in \cite{pg2}. Modified dispersion relation at  energy scales near the Planck energy scale have been used to study black hole radiation in \cite{bhr}. Newtonian white dwarfs composed of degenerate relativistic electron gas describing by modified dispersion relation is studied in, leading to the deformation of mass-radius relation \cite{wd}.

An improvement of the usual  LQC computations of the primordial power spectra  in the inflationary framework was realized in \cite{disploop},  by taking into account the trans-planckian effects by using some modified dispersion relations. In particular, it was found that the unphysical behavior of the deformed algebra spectrum with a quadratic potential and without backreaction can be fixed  by an Unruh-like modified dispersion relation. On the other hand if an exponentially decreasing modified dispersion relation is considered, the nature of the dressed metric spectrum can become unphysical. The consideration of the modified dispersion relations can bring about significant changes in the power spectrum, which, depending on the details of the considered models, may become either compatible or incompatible with the observational data.

The Loop Quantum Cosmology of a flat Friedmann-Lemaître-Robertson-Walker space-time with
a Maxwell type photon field was studied in \cite{lqc}. It turns out  that many of the qualitative properties derived for the case of a massless scalar
field are still valid for the case of  radiation, with the big-bang singularity replaced by a quantum bounce.
Moreover, the operator corresponding to the matter energy density is bounded above by the same critical energy
density.  Evolution equations that very closely approximate the full quantum dynamics
of sharply peaked states at all times were also derived,  and the analytical and
numerical methods used to study other forms of matter fields were improved.

It was suggested in \cite{Miel} that in the limit of high energies loop quantum gravity may lead to the deformation of the local Poincar\'e algebra. The deformation is a direct consequence of the quantum modification of the effective off-shell hypersurface deformation algebra. Interestingly enough, the form of the deformation indicates  that  at large curvatures the signature of the space-time may change from Lorentzian to Euclidean. Particular realizations of the loop-deformed Poincar\'e algebra were also constructed, and it turns out that they can be related to the curved momentum space. By using the corresponding energy dependence of
the group velocity one can obtained some constraints on loop quantum gravity effects  by investigating the time lags of
high energy photons detected from distant astrophysical sources, like, for example,  the gamma ray bursts.

It is the main goal of the present paper to investigate the evolution of the very early Universe in the framework of the Loop Quantum Cosmology by assuming as a matter source a deformed photon gas. When the energy scale approaches the Planck energy  the quantum effects on the very structure of the space-time must also be taken into consideration. One of the important physical implications of the space-time noncommutativity is the modification of the dispersion relation of particles, which has important implications on the statistical mechanics of particles.

In the following we will investigate the statistical physics and LQC implications of
three different models of non-commutative space-times, which lead to modified dispersion relations. The first one has the space-time commutators described by $[x^{i},t]=i\lambda x^{i}$ and $[x^{i},x^{j}]=0$, whose astrophysics properties were studied in \cite{nc1}. The second and the third model are described by Modified Heisenberg-relation $[x_{i},p_{j}]=i\hbar \delta_{ij}[1+\beta p^{2}]$ and $[x_{i},p_{i}]=i\hbar[\delta_{ij} -\alpha (p\delta_{ij} +p_{i}p_{j}/p)+\alpha^{2} (p^{2}\delta_{ij}+3p_{i}p_{j})]$, in which $\beta $ and $\alpha $ are defined by $\beta_{0} /M_{p}^{2}c^{2}$ and $\alpha_{0} /M_{p}c$ \cite{nc2}. The modified Heisenberg relation is related to the non-commutative space via the Jacobi identity.  The dimensionless parameter $\beta_{0} $ and $\alpha_{0} $ can be constrained by many condensed matter experiments, astrophysical experiments and high energy physics experiments \cite{nc2, et1, et2, 87Rb} etc. For example, one of the astrophysical experiment that determines the constraints on $\alpha_{0} $ and $\beta_{0} $ is the gravitational wave event GW150914 \cite{nc2}. As a next step in our study we consider the statistical mechanics of the deformed photon gas satisfying modified energy-momentum dispersion relations. We begin by writing down the general expression of the partition function $Z$ for a thermodynamic system, which allows us to obtain the general expressions of the energy density and pressure of the photon gas satisfying modified dispersion relations. The general formalism allows us to obtain the equations of state of the radiation in the cosmologically relevant limit of high temperatures and densities, or, more exactly, near the Planck scale regimes. The presence of the noncommutativity of the space-time leads to significant modifications in the temperature dependence of the photon gas energy density and pressure, as compared to the standard quantum case of the black body radiation.

At the very early stages of evolution of the Universe, the general relativistic  Friedmann equations are not valid since the quantum effect of gravity must be taken into consideration. Loop Quantum Cosmology, as an application of Loop-Quantum-Gravity, has been widely used to investigate the  Big-Bang singularity. By investigating different LQC models, it was found that the Big-Bang singularity is replaced by a Big Bounce \cite{Asthekar6}-\cite{Brizuela2}, which offers a solution to one of the most important problems in theoretical cosmology.   In the present paper we investigate the LQC-modified Friedmann equations, representing  effective cosmological evolution equation derived from LQG,  together with the assumption that the Universe was initially filled with a photon gas described  the deformed statistical mechanics induced by the noncommutativity of the space-time. We consider explicitly four cases of deformed dispersion relations, and in each case the energy density and pressure of the photon gas are obtained in the high (Planck) temperature limit. Then the modified Friedmann equations are investigated numerically by using the deformed radiation fluid equations of state. The time behaviors of the scale factor, Hubble function, temperature (energy density), and deceleration parameter are considered in detail for different numerical values of the model parameters. The dynamical evolution is studied beginning with an initial contracting phase, in which the scale factor is a monotonically decreasing function of time. After the scale factor reaches a minimum finite value, the Universe enters in an expanding phase. The maximum finite temperature (energy density) is reached at the moment of the bounce, and in the expanding phase the energy density of the photon gas is monotonically decreasing. The behavior of the decelerating parameters indicates that in the contracting phase the Universe did experience a super-accelerated contraction, with the deceleration parameter having high negative values, while during the expanding phase the deceleration parameter reaches positive values, indicating a decelerating behavior.

The present paper is organized as follows.  In Section~\ref{sect2}, we briefly review the  modified dispersion relations to be considered, as well as the general formulation of the deformed quantum statistical mechanics.  We introduce the basic ideas of LQG and  the LQC-modified Friedmann equations in Section~\ref{sect3}, and formulate them  in a dimensionless form.  In Section~\ref{sect4}, the cosmological evolution of the deformed radiation filled very early Universe is investigated for three different photon dispersion relations by using the LQC-modified Friedmann equations. In each case the cosmological evolution equations are integrated numerically, and the time behaviors of the scale factor, of the Hubble function, of the temperature of the Universe, and of the deceleration parameter are studied in detail.  We discuss and conclude our results in Section~\ref{sect5}.  A closed form representation of the equations of state of the deformed photon gas with the dispersion relation $E=kc\sqrt{1-2\beta k^{2}}$ is presented Appendix~\ref{app1}. The invariant phase space volume for the photon dispersion relation  $E=kc(1-\alpha k)$ is calculated in Appendix~\ref{app3}.

\section{Modified dispersion relation from non-commutative space-time, and deformed statistical mechanics}\label{sect2}

The relation between non-commutative nature of the space-time and the modified dispersion relations can be understood in various way. Quantum group methods can be used to find the modified dispersion relation with given commutator of space-time \cite{qg}. Another approach to understand this  relation is via the minimum length implied by the non-commutativity. The four-dimensional momentum operator should be modified since the space-time transformation is only allowed for the multiples of the minimum length. Imposing Lorentz invariance with modified four-dimensional momentum leads to  deformed dispersion relations. Another approach to investigate the relation between non-commutative space-time and modified dispersion relation is using non-commutative geometry \cite{ncg1, ncg2}.

\subsection{Modified dispersion relations}

In this work, three models of non-commutative space-time will be discussed, which lead to three deformed dispersion relations.

\subsubsection{The first non-commutative space-time model}

The first non-commutative space-time model is described by the commutators \cite{nc1}
\begin{equation}
[x^{i},t]=i\lambda x^{i}, [x^{i},x^{j}]=0,
\end{equation}
in which $\lambda$ is a constant parameter.
The dispersion relation of above commutators can be studied by $\kappa$-deformed Poincar\'{e} group, with space-time commutators $[\hat{x_{0},\hat{x_{i}}}]=\frac{i}{\kappa}\hat{x_{i}}$ and $[\hat{x_{l},\hat{x_{j}}}]=0$, which was studied in \cite{vsl}. Using the invariant wave operator
\be
C_{1}^{bcp}\left(1-\frac{C_{1}^{bcp}}{4\kappa^{2}}\right),
\ee
 in the $\kappa$-deformed Minkowski space \cite{vsl}, in which for massless particles,
 \be
 C_{1}^{bcp}=\left(P^{2}\right)\exp^{-P_{0}/\kappa}-2\kappa\sinh\left(\frac{P_{0}}{2\kappa}\right)^{2},
 \ee
 the dispersion relation is obtained in \cite{vsl} as
\begin{equation}
\omega^{2}=\left[\kappa \ln \left(1+\frac{k}{\kappa}\right)\right]^{2}.
\end{equation}
Considering the varying speed of light generalization $E^{2}-p^{2}c^{2}f^{2}=0$ of the above dispersion relation \cite{vsl}, taking $f(E)$ to be $f=1+\lambda E$ \cite{vsl1}, the dispersion relation corresponding to the first non-commutative space-time model considered in this paper can be written as \cite{nc1, vsl1, vsl}
\begin{equation}
E^{2}=k^{2}c^{2}\left( 1+\lambda E\right) ^{2}.
\end{equation}

\subsubsection{The second non-commutative space-time model}

The second non-commutative space-time model is described by the commutators \cite{nc2, mhr1}
\begin{equation}
\begin{aligned}
&[x^{i},p^{j}]=i\hbar \delta^{ij} [1+\beta p^{2}],\\& [p^{i},p^{j}]=0.
\end{aligned}
\end{equation}

The operators for position $x_{i}$ and momentum $p_{i}$ are defined as \cite{nc2}
\begin{equation}
\begin{aligned}
&x_{i}=x_{0i}, \\&p_{i}=p_{0i}(1+\beta p^{2}),
\end{aligned}
\end{equation}
in which $x_{0i}$ and $p_{0i}$ satisfy the canonical commutation relation. Using the relation
\be
p_{a}p^{a}=g_{ab}p^{a}p^{b}=-g_{00}(p^{0})^{2}+g_{ij}p^{0i}p^{0j}(1+\beta p^{2})^{2},
\ee
 and retaining the terms up to $O(\beta ^{2})$ only, the above equation can be written as \cite{nc2}
\begin{equation}
p_{a}p^{a}=-(p^{0})^{2}+p^2+2\beta p^{2}p^{2}=-m^{2}c^{2}+2\beta p^{2}p^{2}.
\end{equation}
Equivalently, the equation can be reformulated as \cite{nc2}
\begin{equation}
(p^{0})^{2}=m^{2}c^{2}+p^{2}(1-2\beta p^{2}).
\end{equation}
Thus we obtain the dispersion relation \cite{nc2}
\begin{equation}
E^{2}=m^{2}c^{2}+p^{2}c^{2}(1-2\beta p^{2}).
\end{equation}

\subsubsection{The third non-commutative space-time model}

The last non-commutative space-time model investigated in this paper is described by the commutators \cite{nc2, mhr2, mhr3}
\begin{equation}
\begin{aligned}
&[x^{i},p^{j}]=i\hbar (\delta^{ij}-\alpha (p\delta^{ij} +p^{i}p^{j}/p)+\alpha^{2} (p^{2}\delta^{ij} +3p^{i}p^{j})), \\&[p^{i},p^{j}]=0.
\end{aligned}
\end{equation}

By using the similar approach as that we used for the second non-commutative space-time model, the deformed dispersion relation for this model can be obtained as \cite{nc2, ali}
\begin{equation}
E^{2}=k^{2}c^{2}(1-\alpha k)^{2}+m^{2}c^{4} .
\end{equation}

\subsection{Deformed quantum statistical mechanics}

In the present work we will investigate the ways the non-commutative space-time influences the cosmological evolution in Loop Quantum Cosmology. In our analysis we will concentrate mostly on the dynamics of a quantum Universe filled with a radiation fluid. As a first step in our study we need to consider the formulation of the quantum statistical mechanics with modified energy-momentum dispersion relations.

According to the basic rules of statistical physics, the grand canonical partition function $Z$ can be written as \cite{landau}
\begin{equation}
Z=\sum_{n_{j}=0}^{\infty}\prod_{j} \left[e^{\left(\mu -E_{j}\right)/k_{B}T}\right]^{n_{j}},
\end{equation}
where $n_j$, $j=1,2,...N$ are the particle numbers, $E_j$ are the corresponding energies of the quantum states, $\mu $ is the chemical potential, while $T$ and $k_B$ denote the temperature, and Boltzmann's constant, respectively.

For radiation, $\mu =0$, and therefore we immediately obtain
\begin{equation}
Z=\prod_{j}\sum_{n_{j}=0}^{\infty} [e^{(-E_{j})/k_{B}T}]^{n_{j}}=\prod_{j} \frac{1}{1-e^{-E_{j}/k_{B}T}},
\end{equation}
and
\begin{equation}
\ln Z=-\sum_{j} \ln \left(1-e^{-E_{j}/k_{B}T}\right),
\end{equation}
respectively.

When the space-time quantum commutators are deformed, the volume of the invariant phase space is also deformed.
We assume that the deformed volume of the invariant phase space can be written as
\be
\frac{d^{3}\vec{k}d^{3}\vec{x}}{f(k,\alpha)},
\ee
 where $\vec{k}$ denotes the particle momentum,  and $\alpha $ is a small parameter in the modified dispersion relation. The function $f\left(\vec{k},\alpha \right)$ must satisfy the condition $\lim _{\alpha \rightarrow 9}f\left(\vec{k},\alpha \right)=1$, a condition ensuring that the invariant phase space volume goes back to $d^{3}\vec{k}d^{3}\vec{x}$ in the limiting case when the dispersion relation takes its standard  form $E=kc$. The energy degeneracy of each state is obtained as
 \be
 \frac{2d^{3}\vec{k}d^{3}\vec{x}}{(2\pi \hbar)^{3}f\left(\vec{k},\alpha\right)},
 \ee
 in which the factor 2 is due to two polarization states of the photon.
In the limit of a continuous system, consisting of identical particles, $\ln Z$ can be written in an integral form as
\begin{align}
&\ln Z=-\int_{0}^{k_{max}} \ln \left(1-e^{-E_{j}/k_{B}T}\right)\frac{2d^{3}\vec{k}d^{3}\vec{x}}{(2\pi \hbar)^{3}f\left(\vec{k},\alpha\right)}\nonumber\\&=-\int_{0}^{k_{max}} \ln \left(1-e^{-E_{j}/k_{B}T}\right)\frac{Vk^{2}dk}{\pi^{2} \hbar^{3}f(k,\alpha)}.
\end{align}
The energy density $u$ of the photon gas can be computed as
\begin{align}
&u=-\frac{1}{V}\partial_{1/k_{B}T}\ln Z\nonumber\nonumber\\&=\int_{0}^{k_{max}}\frac{E}{e^{E/k_{B}T}-1}\frac{k^{2}dk}{\pi^{2}\hbar^{3}f(k,\alpha)}.
\end{align}
The pressure of the gas can be obtained in the form
\begin{align}
&p=k_{B}T\partial_{V}\ln Z\nonumber\\&=-\int_{0}^{k_{max}}\ln \left(1-e^{-E/k_{B}T}\right)\frac{k_{B}Tk^{2}dk}{\pi^{2}\hbar^{3}f(k,\alpha )}\nonumber\\&=-\int_{0}^{k_{max}}\ln \left(1-e^{-E/k_{B}T}\right)\frac{k_{B}Td\left(k^{3}/3\right)}{\pi^{2}\hbar^{3}f(k,\alpha )}\nonumber\\&=\frac{1}{\pi^{2} \hbar^{3}}\int_{0}^{k_{max}}dk\frac{k^{3}}{3f(k,\alpha )\left(e^{E/k_{B}T}-1\right)}\frac{dE}{dk}\nonumber\\&-\frac{k_{B}T}{3\pi^{2} \hbar^{3}}\int_{0}^{k_{max}}\frac{d\left[k^{3}\ln \left(1-e^{-E/k_{B}T}\right)\right]}{f(k,\alpha )}.
\end{align}
Since the last term 
 in the above equation is rather small as compared to the first one, in the following we will neglect it in the expression of the pressure of the deformed radiation fluid. Thus, the basic thermodynamic parameters of a photon gas in the deformed statistical mechanics are obtained as
\begin{equation}\label{u}
\begin{aligned}
u=\frac{1}{\pi^{2}\hbar^{3}}\int_{0}^{k_{max}}dk \frac{k^{2}}{f(k,\alpha )}\frac{E}{e^{E/k_{B}T}-1},
\end{aligned}
\end{equation}
\begin{equation}\label{p}
\begin{aligned}
p=\frac{1}{\pi^{2} \hbar^{3}}\int_{0}^{k_{max}}dk\frac{k^{3}}{3f(k,\alpha )(e^{E/k_{B}T}-1)}\frac{dE}{dk}.
\end{aligned}
\end{equation}

The value of the maximum momentum $k_{max}$, the volume of the invariant phase space $d^{3}\vec{x}d^{3}\vec{k}/f\left(\vec{k},\alpha \right)$, as well as the expression of the dispersion relation $E=E(k)$ are different for the various models of the non-commutative space-time. In the following, by using the expressions (\ref{u}) and (\ref{p}) of the thermodynamic parameters of the photon gas in the deformed statistical mechanics, we will investigate the influence of the non-commutative structure of the space-time on the cosmological evolution of Loop Quantum Cosmological models.

\section{Loop Quantum Cosmology of the deformed radiation gas-the modified
Friedmann equations}\label{sect3}

In a radiation dominated flat FLRW universe, the basic cosmological equations describing the early evolution of the Universe during the quantum gravity dominated phase are the Loop Quantum Cosmology modified Friedmann equations. Loop Quantum Cosmology, which solves the Big-Bang singularity problem by predicting a quantum gravity "bounce", connecting the expanding epoch with the contracting epoch, is an application and an effective theory of Loop Quantum Gravity. The LQC modified Friedmann equations are only qualitatively valid in the early Universe, and they can be applied when the energy scale is comparable to the Planck scale, since in this effective theory
the degrees of freedom with large quantum fluctuations are artificially suppressed.

\subsection{Radiation fields in cosmology}

Before going into Loop Quantum Cosmology, let's first have a look at the classical formalism of cosmology. In the following we assume that the geometry of the Universe can be described by the flat Friedmann-Lemaitre-Robertson-Walker (FLRW) metric, given by
\be
ds^2=-c^2dt^2+a^2(t)\left(dx^2+dy^2+dz^2\right),
\ee
where $a$ is the dimensionless scale factor. We also introduce the Hubble function, defined as $H=\dot{a}/a$.
In such a $k=0$, FLRW universe, the scale factor $\nu $ and its canonical variable $b $ suffice for describing this symmetry-reduced system \cite{lqc}. The definitions of $\nu $ and $b $ are \cite{lqc}
\begin{equation}
\nu=\frac{a^{3}}{\alpha}, b=\frac{\alpha }{2\pi l^{2}_{Pl}}H,
\end{equation}
where $\alpha =2\pi \gamma l^{2}_{Pl}\sqrt{\Delta}$, and  $l_{Pl}$ is the Planck length, defined as $l_{Pl}=\sqrt{G\hbar/c^3}$. $\gamma $ is the Barbero-Immirzi parameter, satisfying the constraint $\ln 2/\pi \leq \gamma \leq \ln 3/\pi $ \cite{gamma1, gamma}. However, in the following for explicit applications we fix the value of $\gamma $ as $\gamma =0.2375$.  $\Delta $ is the smallest nonzero eigenvalue of the LQG area operator, with the order of magnitude $\Delta \approx l_{Pl}^{2}$ \cite{Delta}. The two variables describing the gravitational field in this system satisfy the relations $\{\nu ,b\}=-2/\hbar $ \cite{lqc}. The Hamiltonian constraint density of the gravitational part is given by $-\frac{3\pi G\hbar^{2} }{2\alpha }|\nu |b^{2}$ \cite{lqc}.

In a radiation dominated homogeneous and isotropic Universe, the photon field can be considered as the superposition of three $U(1)$ vector fields, with field strengths given by \cite{lqc}
\begin{equation}
(^{\alpha}A)_{a}=A_{\gamma}(t)\delta^{\alpha}_{a},   (^{\alpha}A)_{t}=0.
\end{equation}

The Hamiltonian constraint density of matter part is given by $\frac{3}{2a}\Pi_{\gamma}^{2}$, where $\Pi_{\gamma}$ is the canonical momentum corresponding to $A_{\gamma}$, satisfying the constraint $\{A_{\gamma},\Pi_{\gamma}\}=\frac{1}{3}$ \cite{lqc} .

The total Hamiltonian constraint is given by the combination of the Hamiltonian constraint of gravitational part and matter part, which is \cite{lqc}
\begin{equation}
NC_{H}=-\frac{3\pi G\hbar^2 }{2\alpha^{2/3} }|\nu |^{4/3}b^{2}+\frac{3}{2}\Pi_{\gamma}^{2}.
\end{equation}
 The Hilbert space of quantized gravitational space-time consists of all those states that can be annihilated by the quantum Hamiltonian constraint.

\subsection{Loop Quantum Cosmology and modified Friedmann equations}

Next we discuss the Loop Quantum Cosmology of the spatially flat, homogeneous and isotropic Universe filled with a radiation fluid.
In the flat FLRW universe dominated by radiation, the kinematic Hilbert space is the product of gravitational and matter Hilbert space \cite{lqc} . Taking $\nu $ as the elementary variable for quantization, the gravitational Hilbert space can be formed by the eigenfunctions of $\hat{\nu } $, which satisfy the condition $<\nu |\nu' >=\delta_{\nu, \nu' }$ \cite{lqc} . For the matter Hilbert space, the elementary variable for quantization is the field $A_{\gamma }$, giving the Hilbert space formed by the eigenfunctions of operator $\hat{A_{\gamma }}$, represented as $|A_{\gamma }>$ \cite{lqc} . Working in the above representation, the Hamiltonian constraint in Loop Quantum Cosmology can be written as \cite{lqc}
\begin{equation}
\hat{NC_{H}}=\Theta \otimes 1+1\otimes \frac{\partial^{2}}{\partial A_{\gamma}^{2}},
\end{equation}
where
\begin{equation}
\begin{aligned}
&\Theta \Psi(\nu; A_{\gamma})=-\frac{9(2\pi \gamma \sqrt{\Delta})^{1/3}}{8\gamma \sqrt{\Delta}\hbar}\nu^{1/3}\times \\&[\theta(\nu -4)(\nu-4)^{1/3}(\nu-2)^{2/3}\Psi(\nu-4;A_{\gamma})\\&-(\theta (\nu -2)(\nu -2)^{2/3}+(\nu +2)^{2/3})\nu^{1/3}\Psi(\nu;A_{\gamma})\\&+(\nu+4)^{1/3}(\nu+2)^{2/3}\Psi(\nu+4;A_{\gamma})].
\end{aligned}
\end{equation}
The physical states are those annihilated by the quantum Hamiltonian constraint. Thus, these states are given by \cite{lqc}
\begin{equation}
-\frac{\partial^{2}}{\partial A_{\gamma}^{2}}\Psi(\nu; A_{\gamma})=\Theta \Psi(\nu; A_{\gamma}).
\end{equation}

It is possible to derive an effective theory of the above LQC in the $k=0$, FLRW radiation dominated universe. The effective Hamiltonian constraint can be written as \cite{lqc}
\begin{equation}
C_{H}^{eff}(t)=-\frac{3\pi G\hbar^{2}}{2\alpha }\nu \sin^{2}b+\frac{3\Pi_{\gamma}^{2}}{2(\alpha \nu )^{1/3}}\approx 0.
\end{equation}
From the effective Hamiltonian constraint we can obtain the Friedmann equation \cite{lqc}
\begin{equation}
\frac{d\nu }{dt}=\frac{6\pi G\hbar }{\alpha }\nu \sin b \cos b,
\end{equation}
which is equivalent to
\begin{equation}\label{lqc1}
H^{2}=\frac{8\pi G }{3}\rho \left(1-\frac{\rho }{\rho_{max} }\right),
\end{equation}
with
\begin{equation}
\rho_{max}=\frac{3c^2}{8\pi G\gamma^{2}\Delta}\approx 1.093\times 10^{94}\times\left(\frac{\gamma ^2}{0.2375}\right)^{-1}\;{\rm g/cm^3}.
\end{equation}
Another equation that must be considered during the cosmological evolution is the energy conservation equation, given by
\begin{equation}\label{lqc2}
\dot{\rho}+3H(\rho +p)=0.
\end{equation}
Taking the time derivative of the Friedmann equation (\ref{lqc1}), and using the energy conservation equation (\ref{lqc2}), it can be easily shown that the time evolution of the Hubble parameter is given by
\begin{equation}\label{lqc3}
\dot{H}=-4\pi G(\rho +p)\left(1-2\frac{\rho}{\rho_{max}}\right).
\end{equation}

\subsubsection{Dimensionless form of the LQC evolution equation}

In order to obtain a dimensionless form of the Friedmann and energy
conservation equations, we introduce a set of dimensionless variables $%
\left(r,P,h,\tau\right)$, defined as
\begin{eqnarray}
r&=&\frac{\rho}{\rho_{max}}, p=\rho_{max}c^2P,  \notag \\
H&=&\sqrt{\frac{8\pi G\rho _{max}}{3}}h, \tau=\sqrt{\frac{8\pi G\rho _{max}}{%
3}}t,
\end{eqnarray}
or, equivalently,
\be
H=\frac{c}{\gamma l_{Pl}}h=\frac{1}{\gamma t_{Pl}}h=7.81\times 10^{43}\times \left(\frac{\gamma }{0.2375}\right)^{-1}h\;{\rm s}^{-1},
\ee
\be
t=\gamma t_{Pl}\tau=1.280\times \frac{\gamma }{0.2375}\times 10^{-44}\;\tau \;{\rm s},
\ee
where the Planck time $t_{Pl}$ is defined as $t_{Pl}=l_{Pl}/c=5.391\times 10^{-44}$ s.

Then the cosmological evolution equations (\ref{lqc1}), (\ref{lqc2}) and (\ref{lqc3}) take the
following dimensionless form,
\begin{equation}  \label{c4}
h^2=r(1-r),
\end{equation}
\begin{equation}  \label{c5}
\frac{dh}{d\tau}=-\frac{3}{2}\left(r+P\right)\left(1-2r\right),
\end{equation}
\begin{equation}  \label{c6}
\frac{dr}{d\tau}+3h\left(r+P\right)=0.
\end{equation}

An important cosmological quantity, the deceleration parameter $q$, is defined according to
\begin{equation}\label{c14}
q=\frac{d}{dt}\frac{1}{H}-1=-\frac{1}{H^{2}}\frac{dH}{dt}-1=-\frac{1}{h^{2}}\frac{dh}{dt}-1.
\end{equation}

In this work, Eqs.~(\ref{c4}), (\ref{c5}) and (\ref{c6}) will be used to investigate the cosmological evolution of the radiation-dominated flat FLRW Universe filled with a deformed radiation fluid, in its early epochs described by the Loop Quantum Cosmology approach, and in the presence of a bounce.

\section{Cosmological evolution of the Loop Quantum Cosmology models with
deformed radiation gas}\label{sect4}

In the present Section we will consider the cosmological evolution of some
classes of Loop Quantum Cosmological models, whose dynamics is described by
the modified Friedmann equations, under the assumption that the matter
content of the early Universe consisted of radiation only. Due to the
quantum nature of the very early Universe, we will also take into account
the quantum effects due to the noncommutative nature of the space-time,
which induces a modification of the energy-momentum dispersion relation for
photons, and, consequently, of the quantum statistical mechanics of the
radiation gas. In our analysis we will investigate the cosmological
implications of three such modifications of the statistical properties of
the photon gas, which are closely connected with the modifications of the
dispersion relations due to the quantum nature of the space-time.

\subsection{LQC implications of the dispersion relation $E=kc\left(1+%
\lambda E\right)$}

As a first example of a cosmological model in LQC with a modified photon gas
we consider the case in which the deformed dispersion relation for photons
is given by
\begin{equation}\label{disp1}
E(k)=kc\left( 1+\lambda E\right) ,
\end{equation}%
where $\lambda $ is a constant. This dispersion relation is particular case
of the more general relation \cite{vsl}
\begin{equation}
E^{2}=k^{2}c^{2}\left( 1+\lambda E\right) ^{2}+m^{2}c^{4},
\end{equation}%
for the case of the massless particles, with $m=0$.

\subsubsection{Thermodynamics of the deformed photon gas}

The energy density of the radiation fluid assumed to fill the Universe when space-time noncommutativity effects are important is given by \cite{nc1}
\begin{equation}
u =\frac{1}{\pi ^{2}\hbar ^{3}c^{3}}\int_{0}^{\infty }\frac{E^{3}}{%
e^{\beta E}-1}\frac{1}{\left( 1+\lambda E\right) ^{4}}dE.  \label{rho1}
\end{equation}

By introducing a new variable $x=\lambda E$, Eq. (\ref{rho1}) can be
rewritten as

\begin{equation}
u =\frac{1}{\pi ^{2}\hbar ^{3}c^{3}\lambda ^{4}}\int_{0}^{\infty }\frac{%
x^{3}}{e^{x/\lambda k_{B}T}-1}\frac{1}{\left( 1+x\right) ^{4}}dx.
\end{equation}

The pressure of the deformed radiation gas is given by \cite{nc1}
\begin{equation}
p=\frac{1}{3\pi ^{2}\hbar ^{3}c^{3}}\int_{0}^{\infty }\frac{E^{3}}{e^{\beta
E}-1}\frac{1}{\left( 1+\lambda E\right) ^{3}}dE,
\end{equation}%
or, equivalently,
\begin{equation}
p=\frac{1}{3\pi ^{2}\hbar ^{3}c^{3}\lambda ^{4}}\int_{0}^{\infty }\frac{x^{3}%
}{e^{x/\lambda k_{B}T}-1}\frac{1}{\left( 1+x\right) ^{3}}dx.
\end{equation}

We will consider the above deformed photon energy density and pressures relations in
two limiting cases.

\paragraph{The case $\lambda k_{B}T>>1$.} First, we assume that the condition $\lambda k_{B}T>>1$ holds.
Then, by expanding the exponential in a power series, and keeping only the
first order approximation, gives

\begin{equation}
u \approx \frac{k_{B}T_{{}}}{\pi ^{2}\hbar ^{3}c^{3}\lambda ^{3}}%
\int_{0}^{\infty }\frac{x^{2}}{\left( 1+x\right) ^{4}}dx=\frac{1}{3}\frac{%
k_{B}T_{{}}}{\pi ^{2}\hbar ^{2}c^{3}\lambda ^{3}}.
\end{equation}

For the pressure of the gas, under the assumption $\lambda k_{B}T>>1$, we obtain
\begin{eqnarray}
p&\approx& \frac{k_{B}T}{3\pi ^{2}\hbar ^{3}c^{3}\lambda ^{3}}%
\int_{0}^{\infty }\frac{x^{2}dx}{\left( 1+x\right) ^{3}}=  \notag \\
&&\frac{k_{B}T}{3\pi ^{2}\hbar ^{3}c^{3}\lambda ^{3}}\int_{0}^{x_{\max }}%
\frac{x^{2}dx}{\left( 1+x\right) ^{3}}=  \notag \\
&&\frac{k_{B}T}{3\pi ^{2}\hbar ^{3}c^{3}\lambda ^{3}}\left[ \ln \left(
1+x_{\max }\right) -\frac{x_{\max }\left( 2+3x_{\max }\right) }{2\left(
1+x_{\max }\right) ^{2}}\right] ,  \notag \\
\end{eqnarray}
where in order to avoid the divergence in the integration we have extended
the limit to the maximum value of $x=\lambda E$ only. By adopting for $%
x_{\max }$ a value of the order of $x_{\max }=\lambda k_{B}T>>1$, we finally
obtain the approximate equation of state of the deformed radiation gas as
given by

\begin{equation}
p\approx \frac{k_{B}T}{3\pi ^{2}\hbar ^{3}c^{3}\lambda ^{3}}\left[ \ln
\left( \lambda k_{B}T\right) -\frac{3}{2}\right] .
\end{equation}

\paragraph {The case $\lambda k_{B}T<<1$.}In the opposite temperature limit, with $\lambda k_{B}T<<1$, we expand $f(x)=x^{3}/\left(
1+x\right) ^{4}$ in Taylor series around zero. Far away from the origin, the
other terms will be exponentially suppressed by the term $\left(
e^{x/\lambda k_{B}T}-1\right) ^{-1}$. By taking into account the formula
\cite{Grad}
\begin{equation}
\int_{0}^{\infty }\frac{x^{\nu -1}dx}{e^{\mu x}-1}=\frac{1}{\mu ^{\nu }}%
\Gamma \left( \nu \right) \zeta \left( \mu \right) ,\mathrm{Re}\;\mu >0,%
\mathrm{Re}\;\nu >0,
\end{equation}%
where $\Gamma \left( \nu \right) $ and $\zeta \left( \mu \right) $ are the
gamma and zeta functions, respectively, we obtain
\begin{eqnarray}
u &\approx& \frac{1}{\pi ^{2}\hbar ^{3}c^{3}\lambda ^{4}}%
\sum_{n=0}^{\infty }\frac{f^{(n)}(0)}{n!}\int_{0}^{\infty }\frac{x^{n}dx}{%
e^{x/\lambda k_{B}T}-1}=  \notag \\
&&\frac{1}{\pi ^{2}\hbar ^{3}c^{3}}\zeta \left( \frac{1}{\lambda k_{B}T}%
\right) \sum_{n=0}^{\infty }\frac{f^{(n)}(0)\Gamma \left( n+1\right) \lambda
^{n-3}}{n!}\times \nonumber\\
&&\left( k_{B}T\right) ^{n+1}.
\end{eqnarray}

For the pressure of the radiation fluid, by taking into account again that  $\lambda k_{B}T<<1$, we easily find
\begin{eqnarray}\label{37}
p&\approx &\frac{1}{3\pi ^{2}\hbar ^{3}c^{3}\lambda ^{4}}\sum_{n=0}^{\infty }%
\frac{g^{(n)}(0)}{n!}\int_{0}^{\infty }\frac{x^{n}dx}{e^{x/\lambda k_{B}T}-1}%
=  \notag \\
&&\frac{1}{3\pi ^{2}\hbar ^{3}c^{3}}\varsigma \left( \frac{1}{\lambda k_{B}T}%
\right) \sum_{n=0}^{\infty }\frac{g^{(n)}(0)\Gamma \left( n+1\right) \lambda
^{n-3}}{n!}\times  \notag \\
&&\left( k_{B}T\right) ^{n+1},
\end{eqnarray}
where $g(x)=x^{3}/\left( 1+x\right) ^{3}$.

\subsubsection{Cosmological dynamics in the case $\lambda k_{B}T>>1$}

We proceed now to the investigation of the cosmological evolution of Loop
Quantum Gravity models with deformed radiation satisfying the dispersion relation (\ref{disp1}).
We will consider first the
case $\lambda k_{B}T>>1$. In this case the energy density and the pressure
of the radiation can be expressed in the dimensionless form
\begin{equation}
r=A\theta ,P=A_1\theta \left( \ln \theta -\frac{3}{2}\right) ,
\end{equation}%
where we have denoted
\begin{equation}\label{39}
\theta =\lambda k_{B}T,A_1=\frac{1}{3\pi ^{2}\hbar ^{3}c^{5}\rho _{max}\lambda
^{4}}.
\end{equation}%
Then Eq. (\ref{c5}) can be written as
\begin{equation}\label{c8}
\frac{dh}{d\tau }=-\frac{3}{2}A_1\theta \left( \ln
\theta -\frac{1}{2}\right) \left(1-2A_1\theta\right).
\end{equation}%

The energy conservation equation (\ref{c6}) takes the form
\begin{equation}\label{c7}
\frac{d\theta }{d\tau }+3h\theta \left( \ln \theta -\frac{1}{2}\right) =0.
\end{equation}

By using Eqs.~(\ref{c14}) and (\ref{c8}), the expression of the deceleration parameter $q$ can be obtained as
\begin{equation}
q=\frac{3}{2}\frac{1} {1-A_1\theta } \left( \ln
\theta -\frac{1}{2}\right)\left(1-2A_1\theta\right)-1.
\end{equation}

To solve the cosmological evolution equations numerically, we need first to estimate the value of the dimensionless parameter $A_1$. In the expression of $A_1$, as given by Eq.~(\ref{39}), $\rho_{max} $ is the maximum density of the Universe, which in the framework of LQC can be obtained as \cite{lqc}
\begin{equation}
\rho_{max} \approx \frac{3}{8\pi G\gamma^{2} l_{Pl}^{2}},
\end{equation}
where the Planck length $l_{Pl}$ is defined as $l_{Pl}=\sqrt{G\hbar/c^3 }$. Substituting the expression of the Planck length, and putting the value of the speed of light back into the expression of $\rho_{max} $, we obtain for $\rho _{max}$ the expression,
\begin{equation}\label{c15}
\rho_{max} =\frac{3c^{5}}{8\pi G^{2}\gamma^{2} \hbar}.
\end{equation}
As for  $A_1$ and $\theta$, they can be written as
\bea
A_1&=&\frac{8G^{2}\gamma^{2} }{9\pi \hbar^{2} c^{10}\lambda^{4} }=7\times 10^{-3}\times \nonumber\\
&&\left(\frac{\gamma }{0.2375}\right)^2\times \left(\frac{\lambda}{10^{-19}\;{\rm GeV}^{-1}}\right)^{-4},
\eea
and
\be
\theta=\lambda E_{Pl}\frac{k_BT}{E_{Pl}}\approx 1.2\times \frac{\lambda }{10^{-19}\;{\rm GeV}^{-1}}\times \frac{T}{T_{Pl}},
\ee
respectively, where $E_{Pl}=k_BT_{Pl}=M_{Pl}c^2\approx 1.2\times 10^{19}$ GeV is the Planck energy, with $M_{Pl}$ denoting the Planck mass $M_{Pl}=\sqrt{\hbar c/G}$. The condition $\lambda k_BT>>1$ can be written as
\be
\lambda E_{Pl}\frac{T}{T_{Pl}}>>1,
\ee
and it can be satisfied in two cases: if $T\approx T_{Pl}$, $\lambda E_{Pl}>>1$, or if   $T>> T_{Pl}$,  $\lambda E_{Pl}\approx 1$, respectively.
Taking $\gamma =0.2375$ \cite{gamma} and $\lambda =10^{-19}\;{\rm GeV}^{-1}$ \cite{lambda}, the numerical value of $A_1$ can be obtained as
$A_1=7\times 10^{-3}$.

\begin{figure*}[htp]
\centering
\includegraphics[width=85 mm]{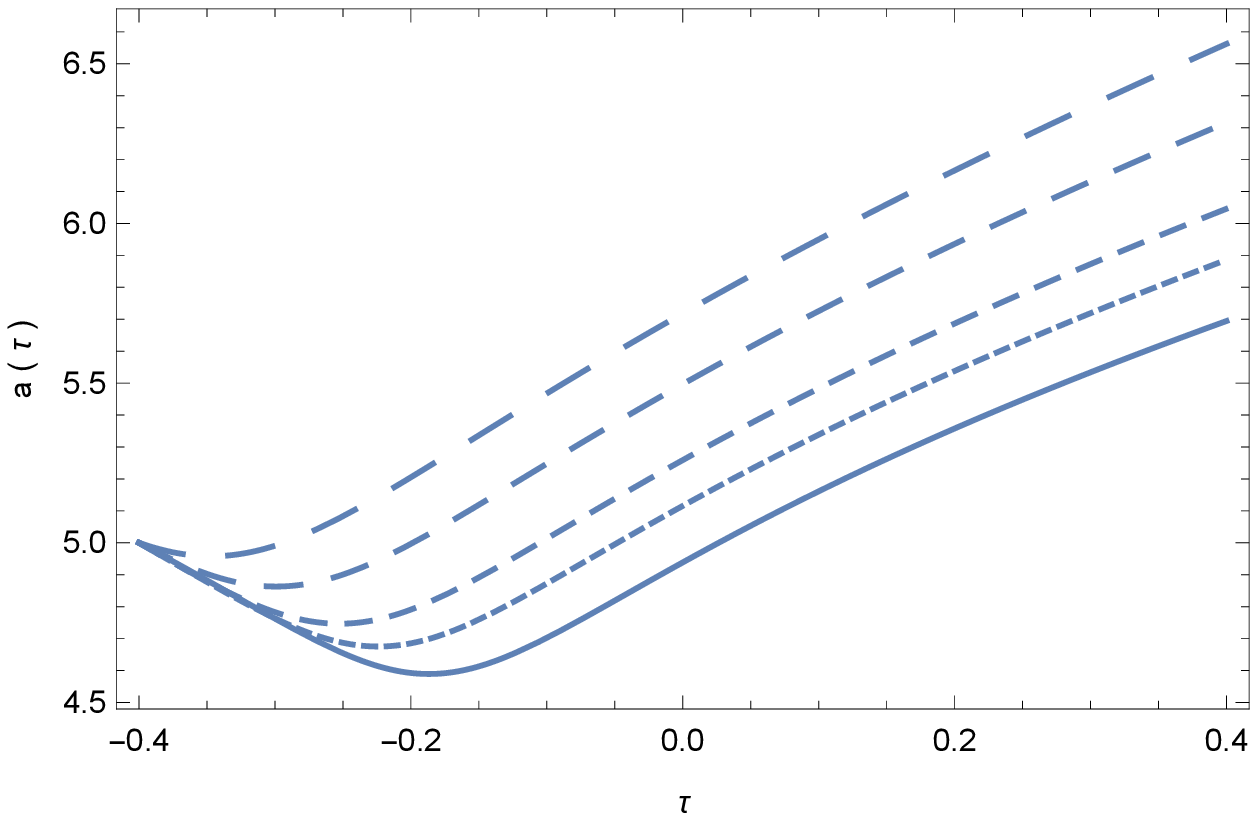}
\includegraphics[width=85 mm]{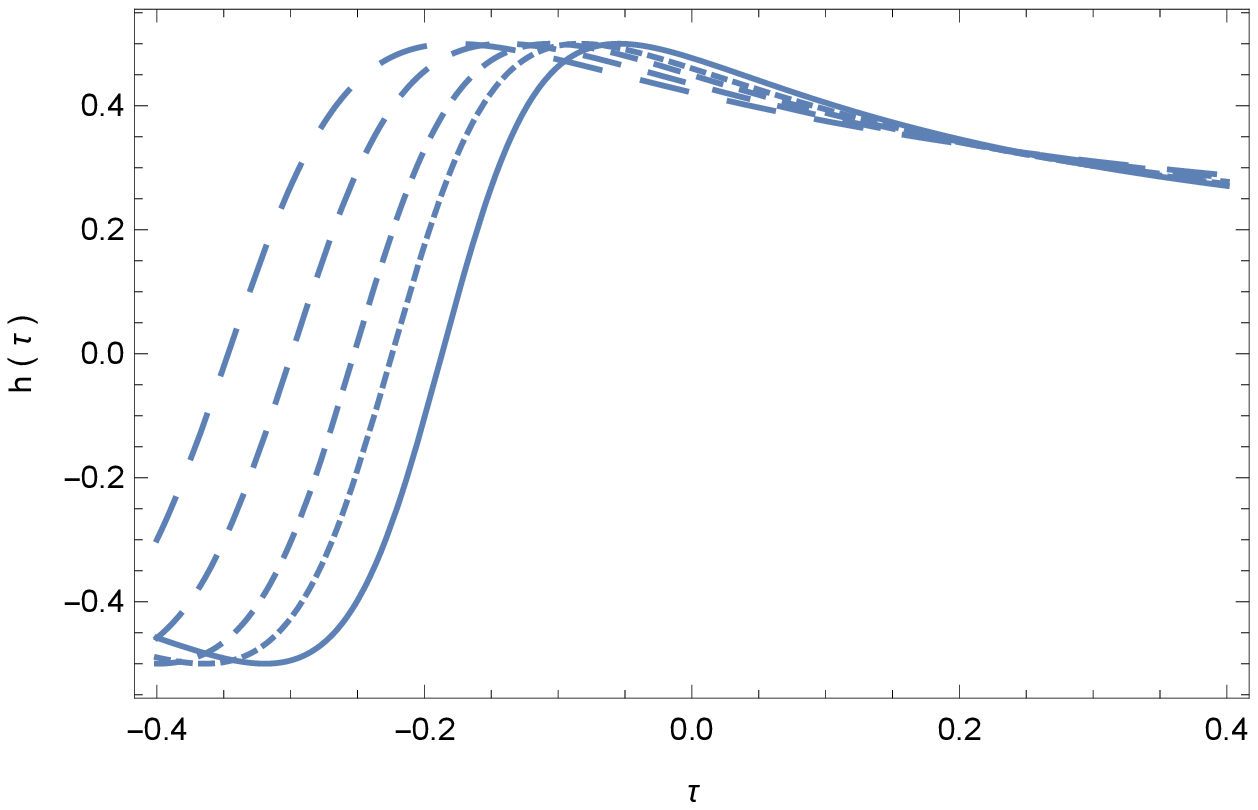}
\caption{Time evolution of the scale factor $a$ (left figure) and of the dimensionless Hubble function $h$ (right figure) in a Loop Quantum Cosmological Universe filled with a deformed photon gas with the dispersion relation $E=kc\left(1+\lambda E\right)$ with  $\lambda k_{B}T>>1$, for different values of the parameter $A_1$: $A_1=3\times 10^{-3}$ (solid curve), $A_1=4\times 10^{-3}$ (dotted curve), $A_1=5\times 10^{-3}$ (short dashed curve), $A_1=7\times 10^{-3}$ (dashed curve), and $A_1=9\times 10^{-3}$ (long dashed curve), respectively. The initial values used to numerically integrate Eqs.~(\ref{c8}) and (\ref{c7}) are $\theta \left(\tau_0\right)=100$, and $h\left(\tau _0\right)=- \sqrt{A_1\theta _0\left(1-A_1\theta _0\right)}$. }\label{fig1}
\end{figure*}

\begin{figure*}[htp]
\centering
\includegraphics[width=85 mm]{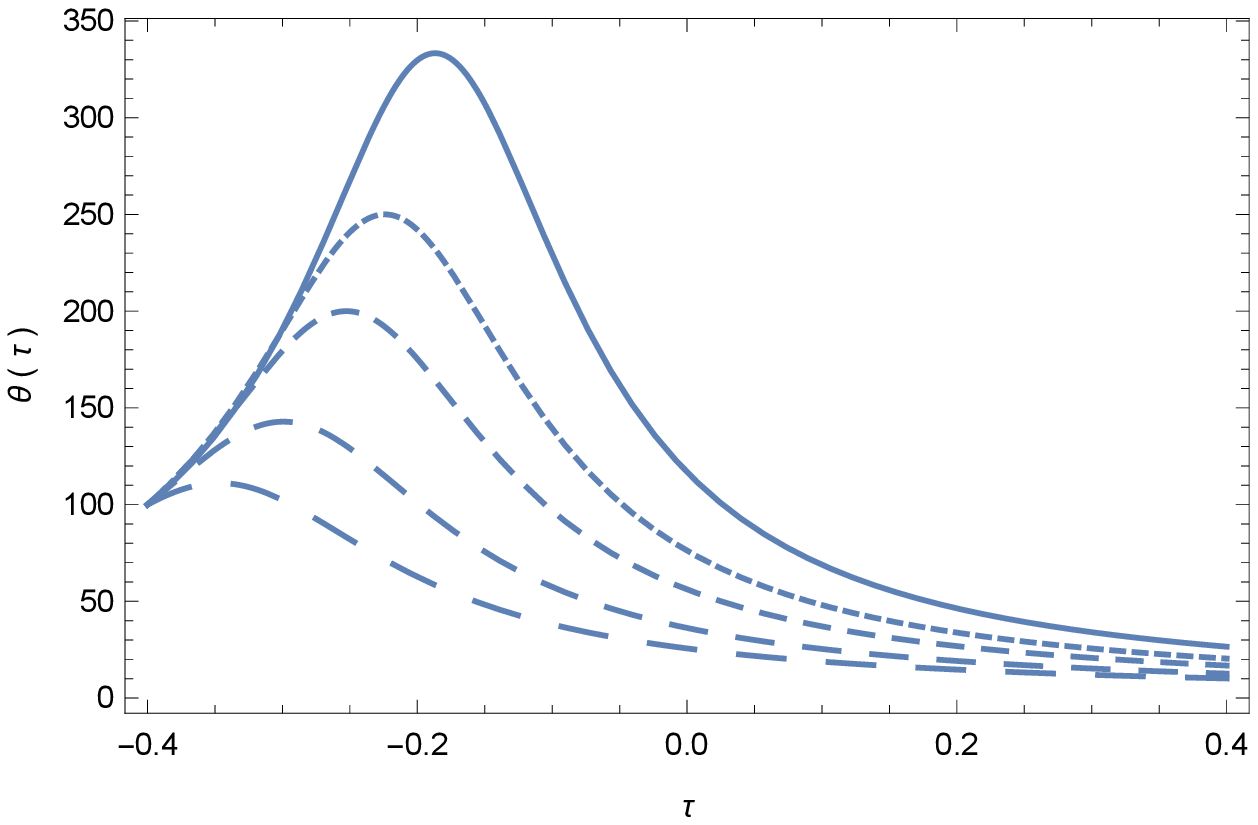}
\includegraphics[width=85 mm]{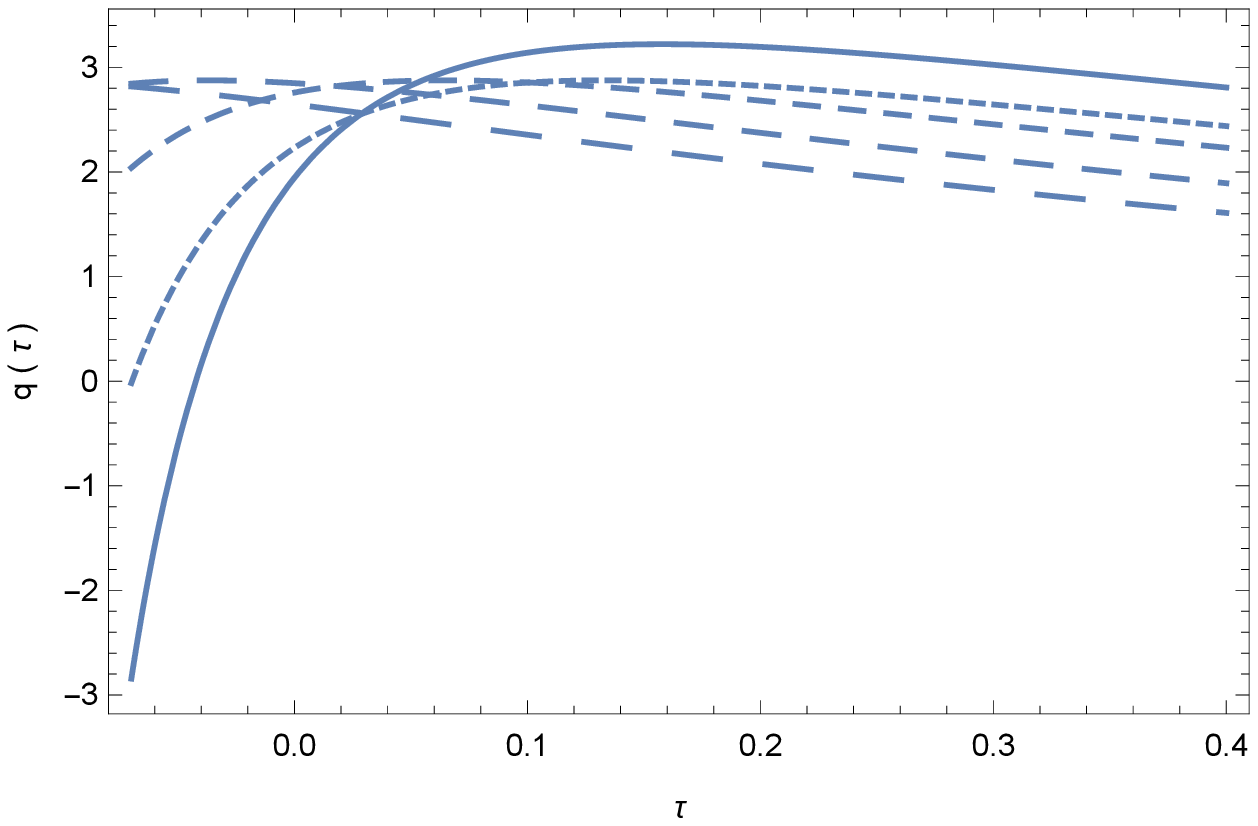}
\caption{Time evolution of the dimensionless temperature $\theta $ (left figure) and of the deceleration parameter $q$ (right figure) in a Loop Quantum Cosmological Universe filled with a deformed photon gas with the dispersion relation $E=kc\left(1+\lambda E\right)$ with  $\lambda k_{B}T>>1$, for different values of the parameter $A_1$: $A_1=3\times 10^{-3}$ (solid curve), $A_1=4\times 10^{-3}$ (dotted curve), $A_1=5\times 10^{-3}$ (short dashed curve), $A_1=7\times 10^{-3}$ (dashed curve), and $A_1=9\times 10^{-3}$ (long dashed curve), respectively. The initial values used to numerically integrate Eqs.~(\ref{c8}) and (\ref{c7}) are $\theta \left(\tau_0\right)=100$, and $h\left(\tau _0\right)= -\sqrt{A_1\theta _0\left(1-A_1\theta _0\right)}$. }\label{fig2}
\end{figure*}

The cosmological evolution of a Universe filled with a deformed radiation fluid satisfying the generalized dispersion relation (\ref{disp1}) and in the limit $\lambda k_BT>>1$ is represented in Figs.~\ref{fig1} and \ref{fig2}, respectively. To obtain the cosmological parameters of interest (scale factor, Hubble function, temperature and deceleration parameter) we have numerically integrated Eqs.~(\ref{c8}) and (\ref{c7}), for different numerical values of the dimensionless parameter $A_1$. In all plots we have fixed the initial value of the temperature $\theta _0$, and obtained the initial value of the dimensionless Hubble function as $h_0=-\sqrt{A_1\theta _0\left(1-A_1\theta _0\right)}$.

As one can see from the left panel of Fig.~\ref{fig1}, the evolution of the Universe begins in a contracting phase, with the scale factor $a$ monotonically decreasing in time. At the time $\tau =\tau _0$ (the bouncing time) the scale factor reaches its minimum finite value, and the Universe bounces into an expanding phase, with the scale factor becoming a monotonically expanding function of time. The cosmological evolution is strongly dependent on the numerical values of the parameter $A_1$, which combines the LQC parameter $\gamma $ and the deformation parameter $\lambda$. The Hubble function, depicted in the right panel of Fig.~\ref{fig1}, shows a complex behavior. Generally, it decreases at the initial stages of the contraction, when it takes negative values. After reaching a local minimum, the Hubble function begins to increase, and becomes zero at the bouncing time, when the scale factor takes its minimum value. After the bounce the Hubble function still continues to increase from the zero to a maximum value, corresponding to a moment $\tau =\tau _{max}$. After that the Hubble function becomes a monotonically decreasing function of time. During the increasing phase, the cosmological evolution is strongly dependent on the numerical values of $A_1$, but in the decreasing phase there is no significant influence of the deformation parameter on the time variation of $h$. The temperature of the Loop Quantum Cosmological Universe, shown in the left panel of Fig.~\ref{fig2}, increasing during the contracting phase, and reaches its maximum at the bounce. After that the temperature of the Universe decreases monotonically. The deceleration parameter $q$, plotted in the left panel of Fig.~\ref{fig2}, indicates a complicated dynamical evolution. During the contracting phase $q$ takes very high negative values, of the order of $q\approx -10^7$, indicating the existence of a super-accelerating phase. However, it increases rapidly, and around the bounce time $\tau _0$ it reaches the marginally accelerating value $q\approx 0$, after which $q$ enters in the region of positive values, indicating a transition to the decelerating phase. After reaching a maximum in the positive domain, the deceleration parameter begins again to decrease. There is a strong influence of the numerical values of the parameter $A_1$ on $q$ in both the contracting and expanding phases.

\subsubsection{The cosmological evolution for $\lambda k_{B}T<<1$}

In the case $\lambda k_{B}T<<1$, the dimensionless equation of state of the deformed radiation fluid is obtained from Eq.~(\ref{37}) as
\begin{equation}\label{c10}
P=\left[\frac{1}{3}+\frac{120\zeta (5)}{\pi^{4} }\lambda k_{B}T\right]r,
\end{equation}
where we have kept only the first two terms in the general series expansion. The dimensionless energy density can be written as
\begin{equation}\label{c11}
r=\frac{k_B^4}{\hbar ^3c^5\rho _{max}}\left[\frac{\pi^2}{15 }-\frac{96\zeta (5)\lambda k_{B}T}{\pi^{2} }\right]T^4.
\end{equation}
By introducing the new variable  $\theta =\lambda k_{B}T$,  and by denoting
\be\label{49}
 A_2=\frac{1}{\lambda^{4} \hbar^{3} c^{5}\rho_{max}}=2\times 10^7\times \left(\frac{\lambda }{10^{-21}\;{\rm GeV}^{-1}}\right)^{-4},
 \ee
Eqs.~(\ref{c11}) and (\ref{c10}) can be written as
\begin{equation}\label{c12}
r=A_2\left[\frac{\pi^2 }{15}-\frac{96\zeta (5)\theta }{\pi^{2} }\right]\theta^{4},
\end{equation}
and
\begin{equation}\label{c13}
P=A_2\left[\frac{\pi^2 }{15}-\frac{96\zeta (5)\theta }{\pi^{2} }\right]\left[\frac{1}{3}+\frac{120\zeta (5)\theta }{\pi^{4} }\right]\theta^{4},
\end{equation}
respectively. Substituting Eqs.~(\ref{c12}) and (\ref{c13}) into Eqs.~(\ref{c5}) and (\ref{c6}), the equation describing the evolution of $h$ and $\theta $ can be obtained as
\bea
\frac{dh}{d\tau }&=&-6A_2\left[ \frac{\pi ^{2}}{15}-\frac{96\zeta (5)\theta }{%
\pi ^{2}}\right] \left[ \frac{1}{3}+\frac{30\zeta (5)\theta }{\pi ^{4}}%
\right] \times \nonumber\\
&&\left[ 1-2A_2\left( \frac{\pi ^{2}}{15}-\frac{96\zeta (5)\theta }{\pi
^{2}}\right) \theta ^{4}\right] \theta ^{4},  \label{mod21}
\eea
and
\begin{equation}
\frac{d\theta }{d\tau }=-\frac{3\left( \frac{\pi ^{2}}{15}-\frac{96\zeta
(5)\theta }{\pi ^{2}}\right) \left[ \frac{1}{3}+\frac{30\zeta (5)\theta }{%
\pi ^{4}}\right] h\theta }{\left( \frac{\pi ^{2}}{15}-\frac{120\zeta
(5)\theta }{\pi ^{2}}\right) },  \label{mod22}
\end{equation}%
respectively. The deceleration parameter, defined by Eq.~(\ref{c14}), is obtained as
\begin{widetext}
\begin{equation}
q=\frac{6A_2\left( \frac{\pi ^{2}}{15}-\frac{96\zeta (5)\theta }{\pi ^{2}}%
\right) \left( \frac{1}{3}+\frac{30\zeta (5)\theta }{\pi ^{4}}\right) \left[
1-2A_2\left( \frac{\pi ^{2}}{15}-\frac{96\zeta (5)\theta }{\pi ^{2}}\right)
\theta ^{4}\right] \theta ^{4}}{A_2\left( \frac{\pi ^{2}}{15}-\frac{96\zeta
(5)\theta }{\pi ^{2}}\right) \left[ 1-A_2\left( \frac{\pi ^{2}}{15}-\frac{%
96\zeta (5)\theta }{\pi ^{2}}\right) \theta ^{4}\right] \theta ^{4}}-1.
\end{equation}
\end{widetext}

Under the above thermodynamics relations, it can be easily shown that only when $A_2\geq 1.03\times 10^{6}$ does the equation  $r-1=0$ have real roots in the real and positive domain. This means that only when $A_2\geq 1.03\times 10^{6}$, the Big Bounce can happen. Equivalently, it means that the upper bound of $\lambda$ is at $\lambda\approx 10^{-21}\; {\rm GeV}^{-1}$. This gives for $\theta $ the numerical expression
\be
\theta \approx 1.2\times 10^{-2}\times \frac{\lambda }{10^{-21}\; {\rm GeV}^{-1}}\times \frac{T}{T_{Pl}}.
\ee

Substituting the expression of $\rho_{max} $ as given by Eq.~(\ref{c15}) back to Eq.~(\ref{49}), and taking $\gamma =0.2375$ and $\lambda =10^{-21}\;{\rm GeV}^{-1}$, we obtain for the numerical value of $A_2$ the estimation $A_2\approx 2\times 10^{7}$.

To solve the cosmological evolution equations numerically, the initial value of $\theta_0$ is also needed. The case being considered now is $\lambda k_{B}T<<1$, or
\be
\lambda E_{Pl}\frac{T}{T_{Pl}}<<1,
\ee
a condition which is satisfied for two choices of the parameters: $T\approx T_{Pl}$, $\lambda E_{Pl}<<1$, or $T<< T_{Pl}$, $\lambda E_{Pl}\approx 1$. For $A_2\approx 2\times 10^{7}$ the condition is equivalent to $T<<1.16\times 10^{34}$ K. Also, when the numerical value of the parameter $A_2$ is of the order $A_2\approx 10^{7}$, the maximum of $\theta$ occurs for values of $\theta \approx 10^{-2}$. Considering that the Planck temperature is $T_{Pl}=1.42\times 10^{32}$ K, the bounce occurs at the temperature scale of Planck temperature when the parameter $A_2$ is of the order of $A_2\approx 10^{7}$. From the above considerations, the initial value of $\theta_0$ in the contracting epoch is taken to be 0.009.

\begin{figure*}[htp]
\centering
\includegraphics[width=85 mm]{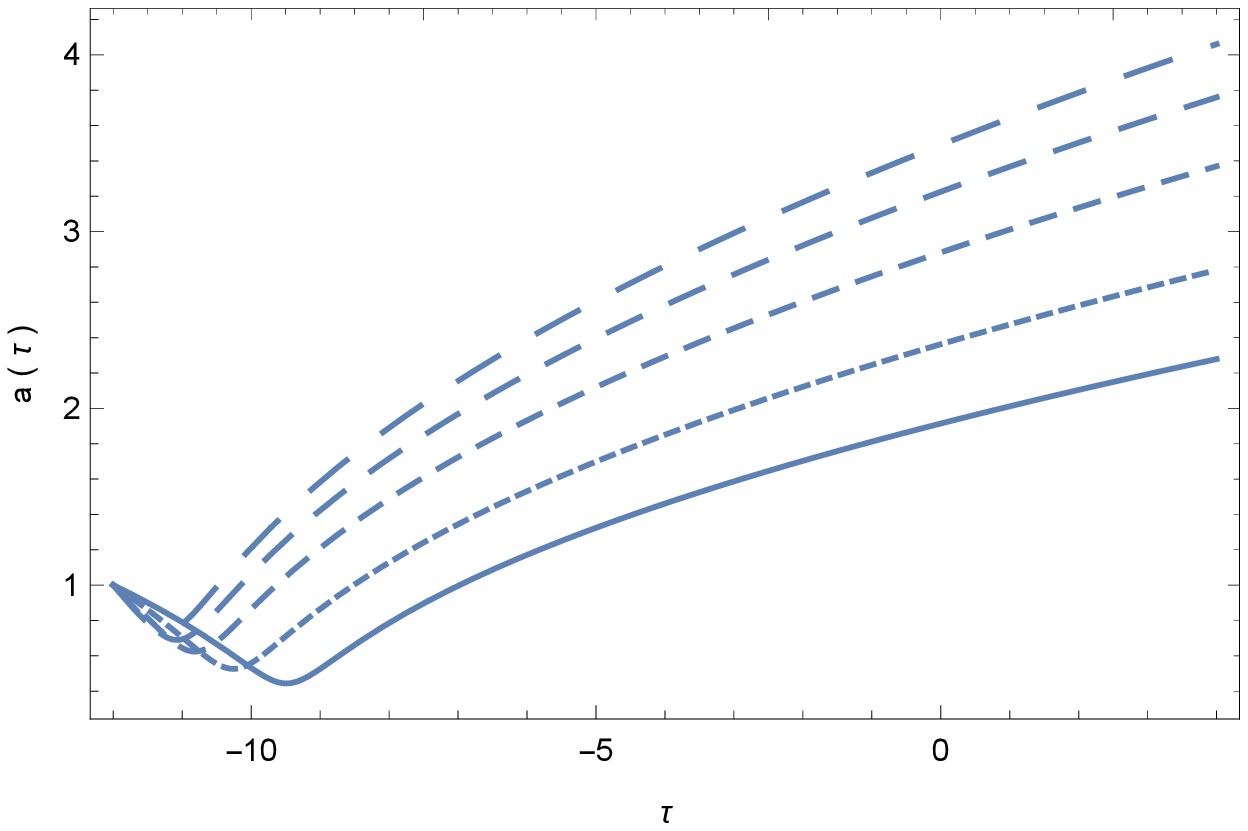}
\includegraphics[width=85 mm]{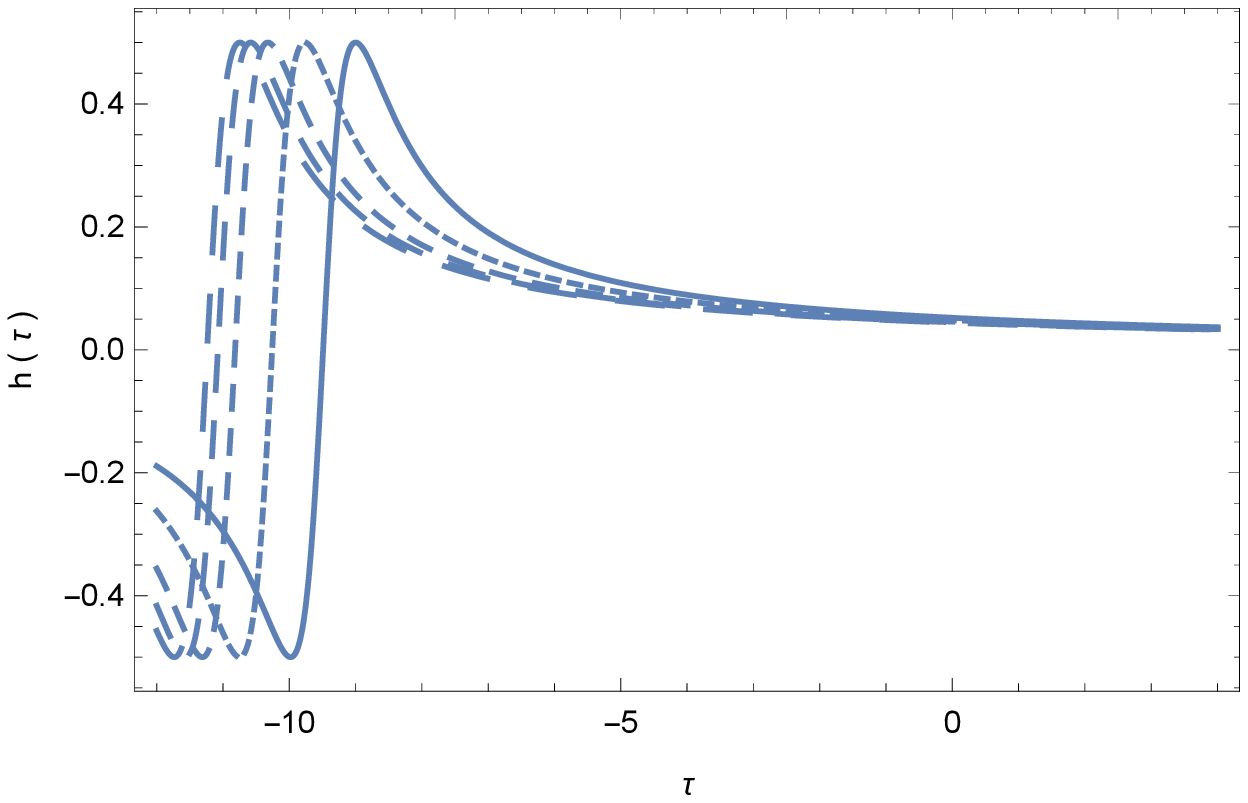}
\caption{Time evolution of the scale factor $a$ (left figure) and of the dimensionless Hubble function $h$ (right figure) in a Loop Quantum Cosmological Universe filled with a deformed photon gas with the dispersion relation $E=kc\left(1+\lambda E\right)$, with  $\lambda k_{B}T<<1$, for different values of the parameter $A_2$: $A_2=10^{7}$ (solid curve), $A_2=2\times 10^{7}$ (dotted curve), $A_2=3\times 10^{7}$ (short dashed curve), $A_2=4\times 10^{7}$ (dashed curve), and $A_2=5\times 10^{7}$ (long dashed curve), respectively. The initial conditions used to numerically integrate Eqs.~(\ref{mod21}) and (\ref{mod22}) are $\theta \left(\tau_0\right)=0.009$,  while $h_0$ is obtained from Eq.~(\ref{69}). }\label{fig3}
\end{figure*}

\begin{figure*}[htp]
\centering
\includegraphics[width=85 mm]{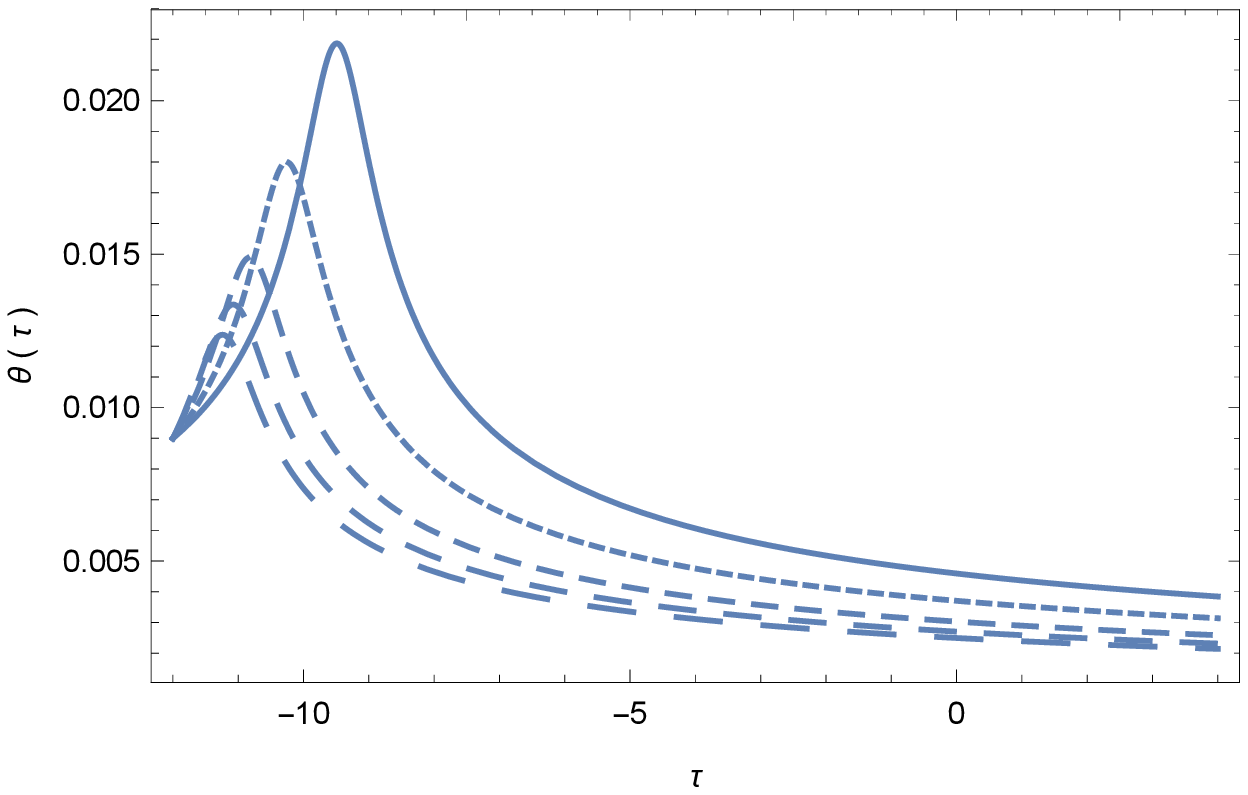}
\includegraphics[width=85 mm]{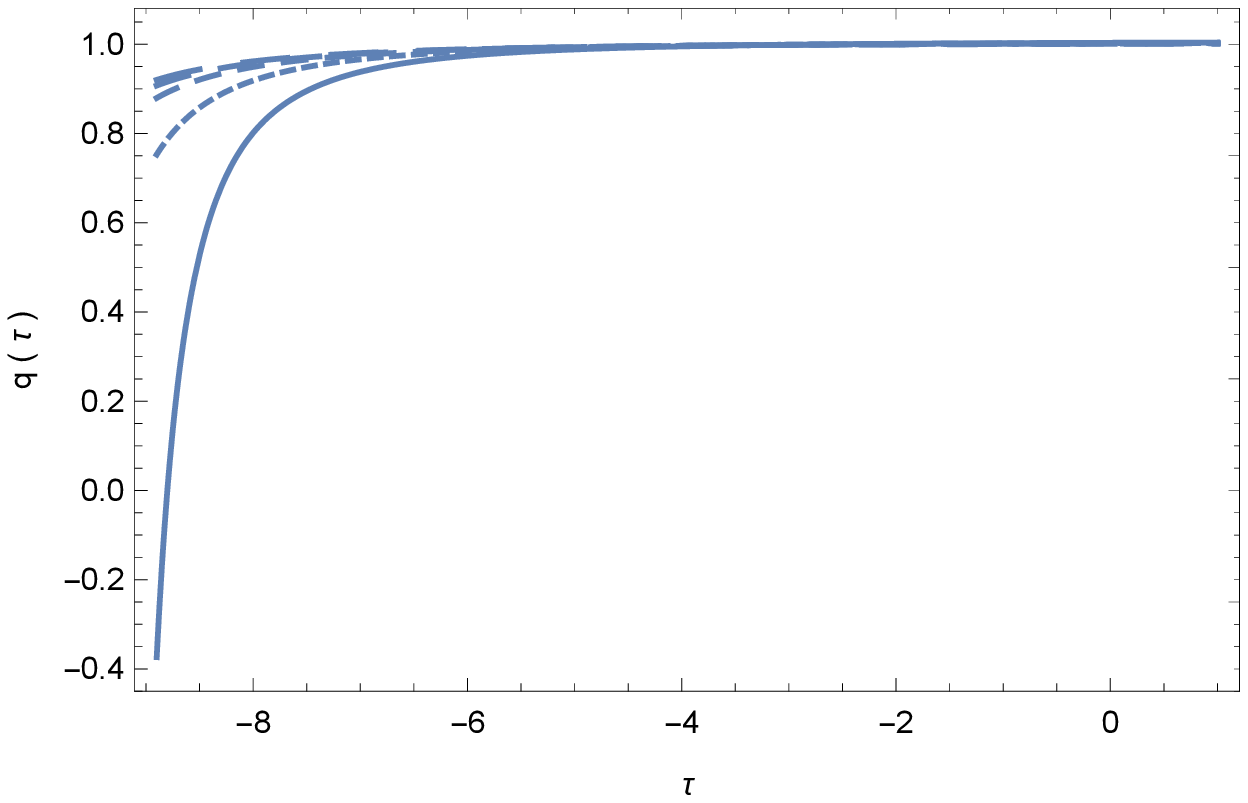}
\caption{Time evolution of the dimensionless temperature $\theta $ (left figure) and of the deceleration parameter $q$ (right figure) in a Loop Quantum Cosmological Universe filled with a deformed photon gas with the dispersion relation $E=kc\left(1+\lambda E\right)$ with  $\lambda k_{B}T<<1$, for different values of the parameter $A_2$: $A_2=10^{7}$ (solid curve), $A_2=2\times 10^{7}$ (dotted curve), $A_2=3\times 10^{7}$ (short dashed curve), $A_2=4\times 10^{7}$ (dashed curve), and $A_2=5\times 10^{7}$ (long dashed curve), respectively. The initial values used to numerically integrate Eqs.~(\ref{mod21}) and (\ref{mod22}) are $\theta \left(\tau_0\right)=0.009$,  while $h_0$ is obtained from Eq.~(\ref{69}). }\label{fig4}
\end{figure*}

The time evolution of the scale factor $a$, of the dimensionless Hubble function $h$, of the temperature of the Universe $\theta $, and of the deceleration parameter $q$ are presented in Figs.~\ref{fig3} and \ref{fig4}, respectively. They are obtained by numerically integrating Eqs.~(\ref{mod21}) and (\ref{mod22}) for different values of the model parameter $A_2$. In order to integrate the evolution equations we fix the value of the initial temperature $\theta \left(\tau _0\right)=\theta _0$, and we obtain the initial value $h\left(\tau _0\right)=h_0$ of the dimensionless Hubble function in the contracting phase as
\bea\label{69}
\hspace{-1.0cm}&&h_0=\nonumber\\
\hspace{-1.0cm}&&-\sqrt{A_2 \theta _0^4 \left(\frac{\pi ^2}{15}-\frac{96 \theta _0 \zeta
   (5)}{\pi ^2}\right) \left[1-A_2 \theta _0^4 \left(\frac{\pi ^2}{15}-\frac{96
   \theta _0 \zeta (5)}{\pi ^2}\right)\right]}.\nonumber\\
\eea

As one can see from Figs.~\ref{fig3} and \ref{fig4}, the time evolution of the deformed radiation filled Universe with $\lambda k_BT<<1$ is qualitatively similar to the opposite temperature limit, considered previously. The Universe begins its evolution in a contracting phase, with the scale factor, shown in the left panel of Fig.~\ref{fig3} decreasing rapidly in time. At a moment $\tau =\tau _0$, the scale factor reaches a minimum value, and the Universe bounces to an expanding phase. The Hubble function $h$, presented in the right panel of Fig.~\ref{fig3}, decreases during the contracting phase, and reaches a local minimum at the bounce time  $\tau _0$, after which the decreasing phase is followed by an increasing one, in which $h$ enters the positive value range, and it reaches a local maximum at a finite moment $\tau _{max}$. After reaching its maximum value, the Hubble function monotonically decreases in time. The temperature of the Universe, depicted in the left panel of Fig.~\ref{fig4}, decreases rapidly during the contracting phase, and reaches its maximum at the bounce time, and then it decreases monotonically. The deceleration parameter, represented in the right panel of Fig.~\ref{fig4},  takes very large negative values during the contracting phase, and reaches the region of positive values approximately at the moment when the Hubble function reaches its maximum value. During the expanding phase that follows after $h$ has reached its maximum, the deceleration parameter is approximately constant, with values of the order of $q\approx 1$, indicating a decelerating expansion of the Universe. Generally, in the early contracting and immediately post contracting phases, the cosmological evolution is strongly dependent on the numerical values of $A_2$. However, for large times, the dynamics of $h$ and $q$ is independent on $A_2$, but the scale factor and temperature evolutions still depend on $A_2$.

\subsection{LQC implications of the dispersion relation $E=kc\sqrt{1-2\beta k^{2}} $}

As the second example of a cosmological model in LQC with a modified photon gas
we consider the case in which the deformed dispersion relation for photons
is given by
\begin{equation}
E=kc\sqrt{1-2\beta k^{2}}  ,
\end{equation}%
where $\beta $ is a constant. This dispersion relation is particular case
of the more general relation \cite{nc2}
\begin{equation}
E^{2}=k^{2}c^{2}(1-2\beta k^{2})+m^{2}c^{4}.
\end{equation}%

\subsubsection{Thermodynamics of the photon gas}

For the case of the massless particles, with $m=0$, the energy density of
the photon gas is given by
\begin{equation}
u=\frac{1}{\pi ^{2}\hbar ^{3}}\int_{0}^{\frac{1}{2\sqrt{\beta }}}\frac{k^{2}E}{
(1+\beta k^{2})^{3}(e^{\beta E}-1)}dk.
\end{equation}

There are two things one should be mentioned about this integral expression: first, the classical degree of degeneracy of each phase unit $\frac{d^{3}\vec{x}d^{3}\vec{k}}{\hbar^{3}}$ is modified to $\frac{d^{3}\vec{x}d^{3}\vec{k}}{(1+\beta \vec{k}^{2})^{3}\hbar^{3}}$ \cite{ips}, due to the modification of the invariant phase space; secondly, the upper limit of integration is taken to be $1/2\sqrt{\beta }$. When $k=1/(2\sqrt{\beta })$, $E=E_{max}$; when $k=1/\sqrt{2\beta }$, $E=0$. Which should be the maximum value of the momentum, $k=1/(2\sqrt{\beta })$ or $k=1/\sqrt{2\beta }$ ?
The correct upper bound for the photon momentum is $k=1/(2\sqrt{\beta })$. If the upper bound of the momentum is taken as $k=1/\sqrt{2\beta }$, the pressure will become negative, which is not physical.

By introducing a new variable $x=kc/k_{B}T$, and denoting
\be
\tilde{T}=\frac{\beta k_{B}^{2}T^{2}}{c^{2}},
\ee
the energy density can be rewritten as
\bea
u&=&\frac{k_{B}^{4}T^{4}}{\pi ^{2}\hbar ^{3} c^{3}}\int_{0}^{\frac{1}{2\sqrt{\tilde{T}}}}\frac{x^{3}\sqrt{1-2\tilde{T}x^{2}}}{
\left(1+\tilde{T}x^{2}\right)^{3}\left(e^{x\sqrt{1-2\tilde{T}x^{2}}}-1\right)}dx \nonumber\\
&&=\frac{k_{B}^{4}T^{4}}{\pi ^{2}\hbar ^{3} c^{3}}I\left(\tilde{T}\right)=\frac{c}{\pi ^2\beta ^2\hbar ^3}\tilde{T}^2I\left(\tilde{T}\right),
\eea
where we have denoted
\begin{equation}\label{75}
I\left(\tilde{T}\right)=\int_{0}^{\frac{1}{2\sqrt{\tilde{T}}}}\frac{x^{3}\sqrt{1-2\tilde{T}x^{2}}}{
\left(1+\tilde{T}x^{2}\right)^{3}\left(e^{x\sqrt{1-2\tilde{T}x^{2}}}-1\right)}dx.
\end{equation}

The pressure of the deformed radiation gas is given by
\begin{equation}
p=\frac{1}{3\pi ^{2}\hbar ^{3}}\int_{0}^{\frac{1}{2\sqrt{\beta }}}dk \frac{k^{3}}{(1+\beta k^{2})^{3}(e^{\frac{E}{k_{B}T}}-1)}\frac{dE}{dk}.
\end{equation}%
or, equivalently, by
\be
p=\frac{c}{3\pi ^2\beta ^2\hbar ^3}\left[I\left(\tilde{T}\right)-2\tilde{T}J\left(\tilde{T}\right)\right]\tilde{T}^2=\frac{c}{3\pi ^2\beta ^2\hbar ^3}K\left(\tilde{T}\right),
\ee
where we have denoted
\bea
\hspace{-0.5cm}&&J\left(\tilde{T}\right)=\nonumber\\
\hspace{-0.5cm}&&\int_{0}^{\frac{1}{2\sqrt{\tilde{T}}}}\frac{x^{5}}{\left(1+\tilde{T}x^{2}\right)^{3}\sqrt{1-2\tilde{T}x^{2}}(e^{x\sqrt{1-2\tilde{T}x^{2}}}-1)}dx.
\eea

\begin{figure*}[htp]
\centering
\includegraphics[width=85 mm]{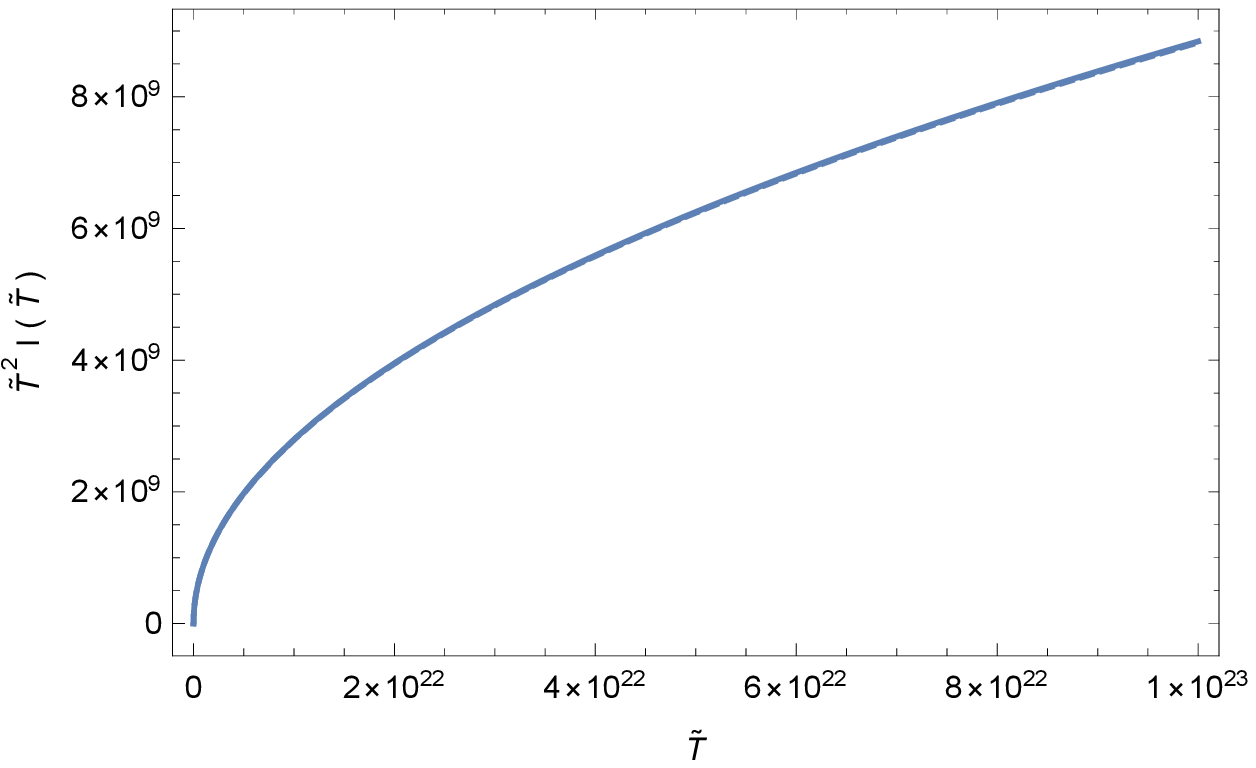}
\includegraphics[width=85 mm]{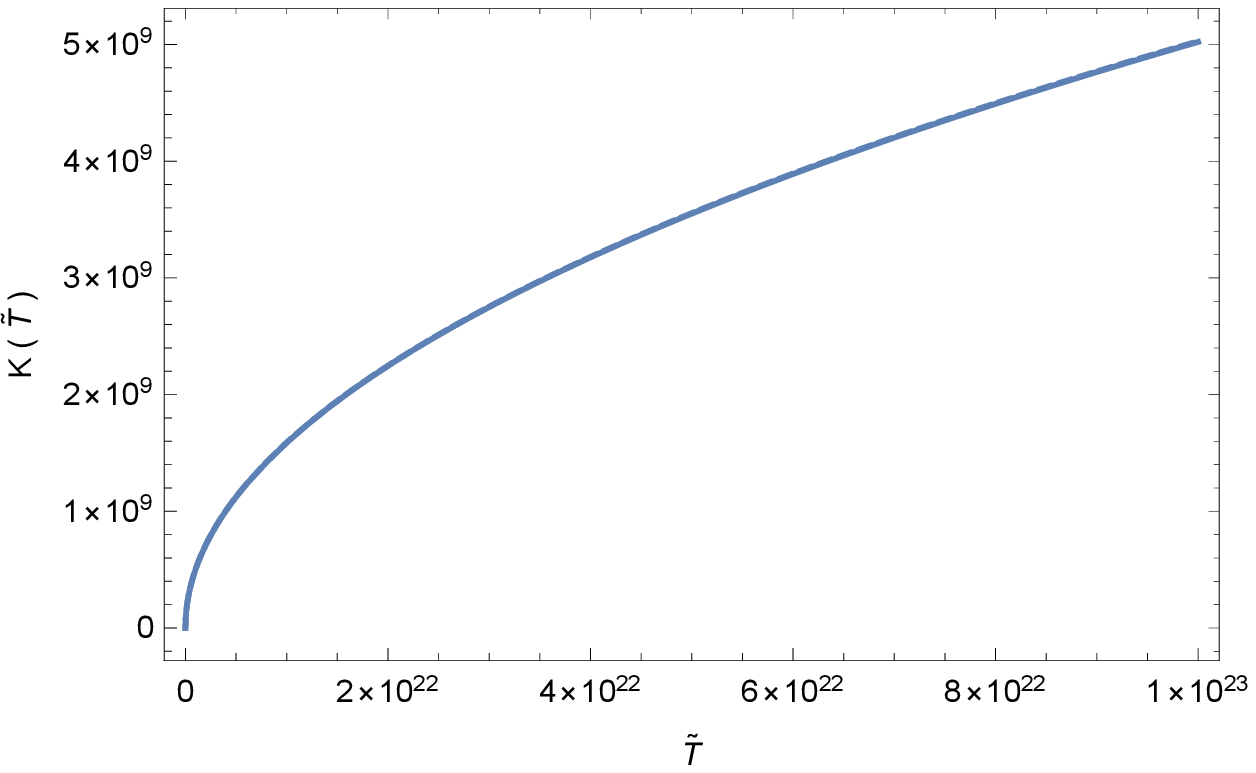}
\caption{Variations of the functions $\tilde{T}^2I\left(\tilde{T}\right)$ (solid curve) and $0.0279\sqrt{\tilde{T}}$ (dotted curve) (left figure) and of the functions $K\left(\tilde{T}\right)$ and  $0.0159\sqrt{\tilde{T}}$ (dotted curve) (right figure).}\label{fig5}
\end{figure*}

The variations of the function $\tilde{T}^2I\left(\tilde{T}\right)$ and $K\left(\tilde{T}\right)$ are represented in Fig.~\ref{fig5}.

The numerical results can be fitted with a high precision by two simple functions, proportional to $\sqrt{\tilde{T}}$. Hence we can represent the energy density and the pressure of the photon gas as
\be
u=\frac{0.0279}{\pi^2\beta ^2\hbar ^3}\sqrt{\beta}k_BT,
\ee
and
\be
p=\frac{0.0159}{3\pi^2\beta ^2\hbar ^3}\sqrt{\beta}k_BT,
\ee
respectively. Exact closed form analytical representations of the energy density and pressure of the photon gas satisfying the modified dispersion relation $E=kc\sqrt{1-2\beta k^{2}} $ are presented in Appendix~\ref{app1}.

\subsubsection{Loop Quantum Cosmological evolution of the deformed radiation filled Universe with $E=kc\sqrt{1-2\beta k^{2}} $}

We proceed now to the investigation of the cosmological evolution of Loop
Quantum Gravity models with deformed radiation described by the modified dispersion relation $E=kc\sqrt{1-2\beta k^{2}} $. In this case the energy density and the pressure
of the radiation can be expressed in the dimensionless form
\begin{equation}\label{c16}
r=0.02796 A_{3} \theta,\,\,
P=\frac{0.01588 A_{3} \theta}{3},
\end{equation}%
where we have denoted
\begin{equation}
\theta =\frac{\sqrt{\beta }k_{B}T}{c},\,\,A_{3}=\frac{1}{\pi ^{2}\hbar ^{3}\beta^{2} c\rho _{max}}.
\end{equation}%
Then Eq. (\ref{c5}) can be written as
\begin{equation}\label{c8}
\frac{dh}{d\tau }=-0.04988A_3\theta \left(1-0.05592 A_{3} \theta\right).
\end{equation}%

The energy conservation equation (\ref{c6}) takes the form
\begin{equation}\label{c7}
\frac{d\theta }{d\tau }+3.5679h\theta =0.
\end{equation}

By using Eqs.~(\ref{c14}) and (\ref{c8}), the expression of the deceleration parameter $q$ can be obtained as
\begin{equation}
q=\frac{1.7893\left(1-0.05592 A_{3} \theta \right)}{ 1-0.02796 A_{3} \theta}-1.
\end{equation}

To solve the cosmological evolution equations (\ref{c8}) and (\ref{c7})  numerically, we need to obtain first the value of $A_{3}$.
Substituting the expression of $\rho_{max}$ as given by Eq.~(\ref{c15}) back into $A_{3}$, we find
\begin{equation}
A_{3}=\frac{8G^{2}\gamma^{2} }{3\pi \hbar^{2} c^{6}\beta^{2} },
\end{equation}
which is a dimensionless quantity.
In the expression of $A_{3}$, $\beta $ can be written as $\beta_{0} /M_{Pl}^{2}c^{2}$, where $M_{Pl}$ is the Planck mass \cite{nc2}. The Planck energy $M_{Pl}c^{2}$ is given by $1.2\times 10^{28}\;{\rm eV}$. For the dimensionless quantity $\theta $ we find
\be
\theta =\frac{\sqrt{\beta _0}}{M_{Pl}c^2}E_{Pl}\frac{T}{T_{Pl}}=\sqrt{\beta _0}\frac{T}{T_{Pl}}.
\ee
For $\gamma =0.2375$ and $\beta_{0} =10^{21}$ (the upper bound of $\beta_{0} $ is obtained from electron tunnelling experiments) \cite{et1, et2}, the numerical value of $A_{3}$ can be obtained  approximately as
\begin{equation}
A_{3}=4.45\times 10^{-44}.
\end{equation}

To investigate the cosmological evolution, the initial value of $\theta $ is needed. In the contracting epoch, the initial temperature is taken to be $0.1T_{Pl}$, corresponding to $\theta \left(\tau_0\right)=3.22\times 10^{9}$. In the expanding epoch, the initial temperature (the temperature of the bounce) is taken to be the Planck temperature $T_{Pl}=1.42\times 10^{32}$ K, corresponding to $\theta \left(\tau_0\right)=3.22\times 10^{10}$. Hence we investigate the evolution of the Universe within the Loop Quantum Cosmological framework with a modified radiation dispersion relation in the range of temperatures of $0.1 T_{Pl}<T<T_{Pl}$. The initial value of the Hubble function $h\left(\tau _0\right)=h_0$ is determined by the initial temperature $\theta \left(\tau _0\right)=\theta _0$ of the Universe as
\be\label{87}
h_0=-\sqrt{0.02796 A_{3} \theta_0\left(1-0.02796 A_{3} \theta_0\right)}.
\ee

\begin{figure*}[htp]
\centering
\includegraphics[width=85 mm]{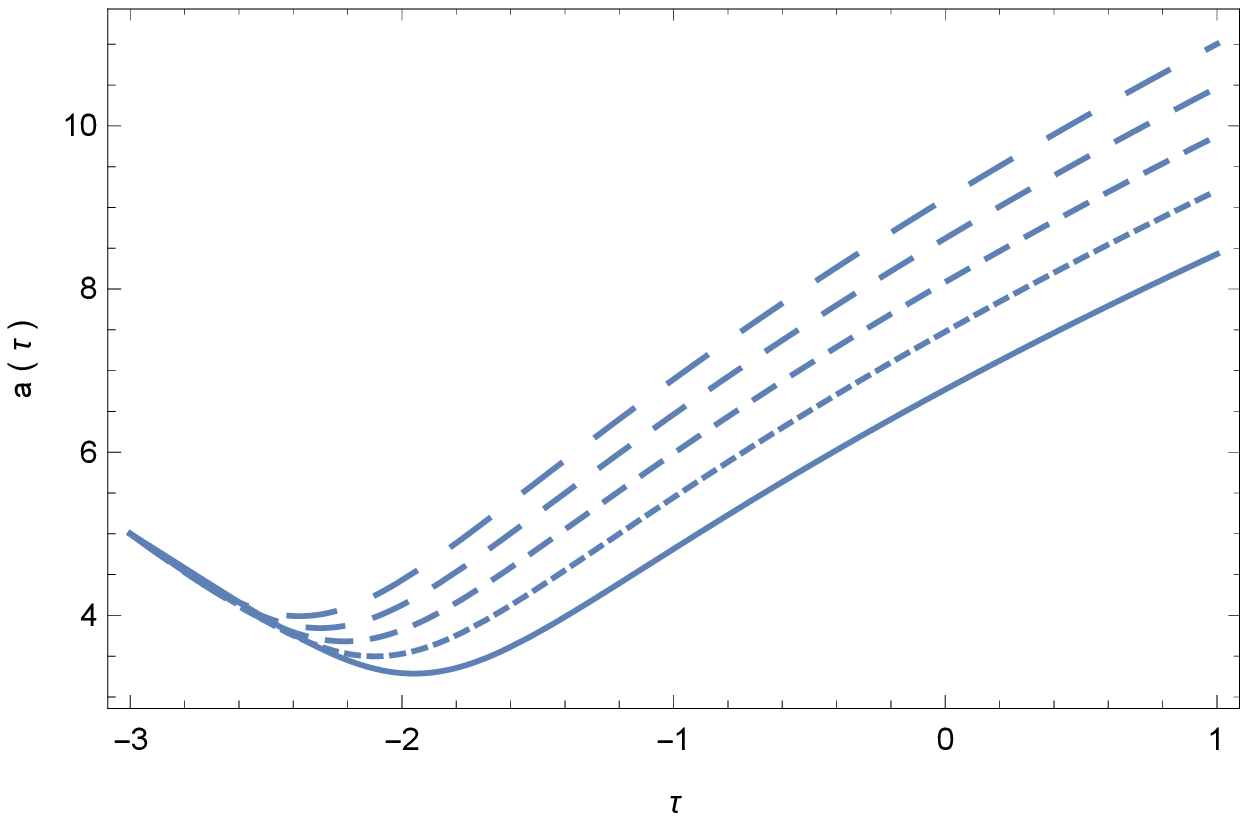}
\includegraphics[width=85 mm]{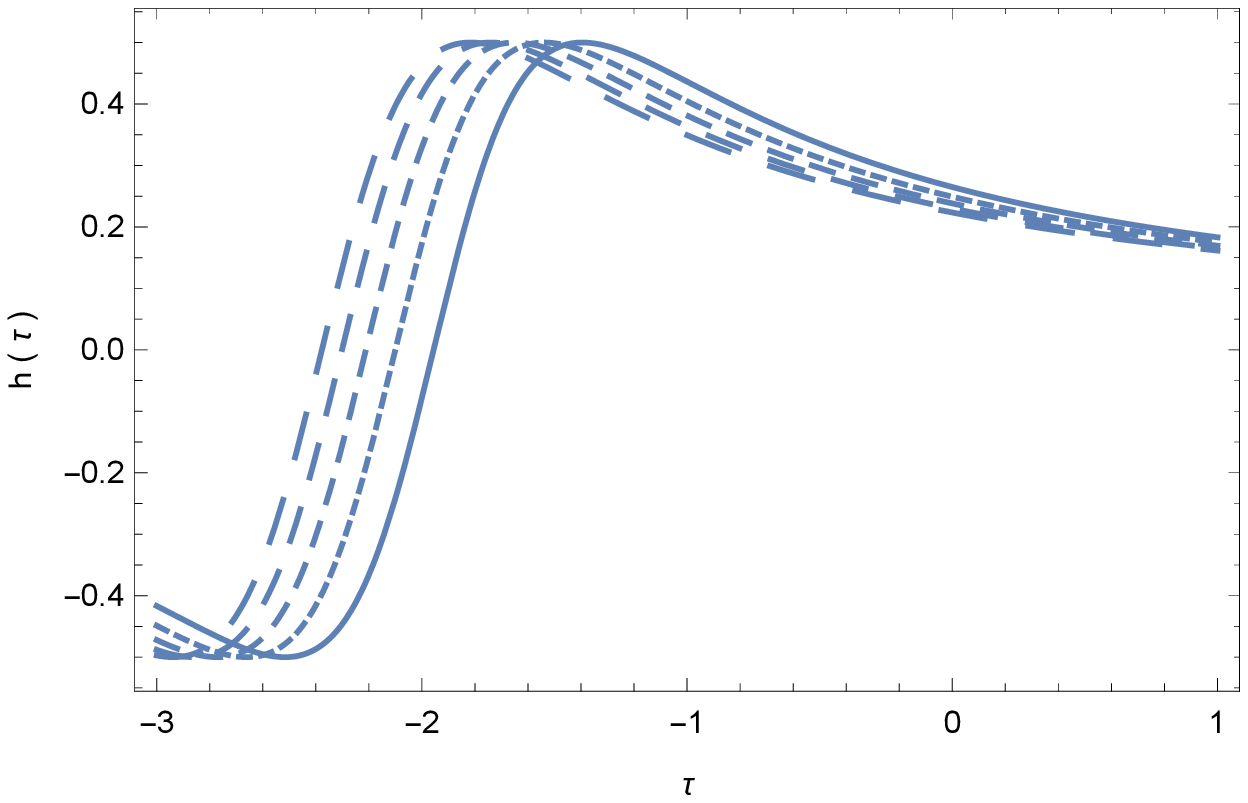}
\caption{Time evolution of the scale factor $a$ (left figure) and of the dimensionless Hubble function $h$ (right figure) in a Loop Quantum Cosmological Universe filled with a deformed photon gas satisfying the dispersion relation $E=kc\sqrt{1-2\beta k^{2}} $, for different values of the parameter $A_3$: $A_3=4\times 10^{-44}$ (solid curve), $A_3=5\times 10^{-44}$ (dotted curve), $A_3=6\times 10^{-44}$ (short dashed curve), $A_3=7\times 10^{-44}$ (dashed curve), and $A_3=8\times 10^{-44}$ (long dashed curve), respectively. The initial conditions used to numerically integrate Eqs.~(\ref{mod21}) and (\ref{mod22}) are $\theta \left(\tau_0\right)=2\times 10^{44}$, and with $h_0$ obtained from Eq.~(\ref{87}). }\label{fig6}
\end{figure*}

\begin{figure*}[htp]
\centering
\includegraphics[width=85 mm]{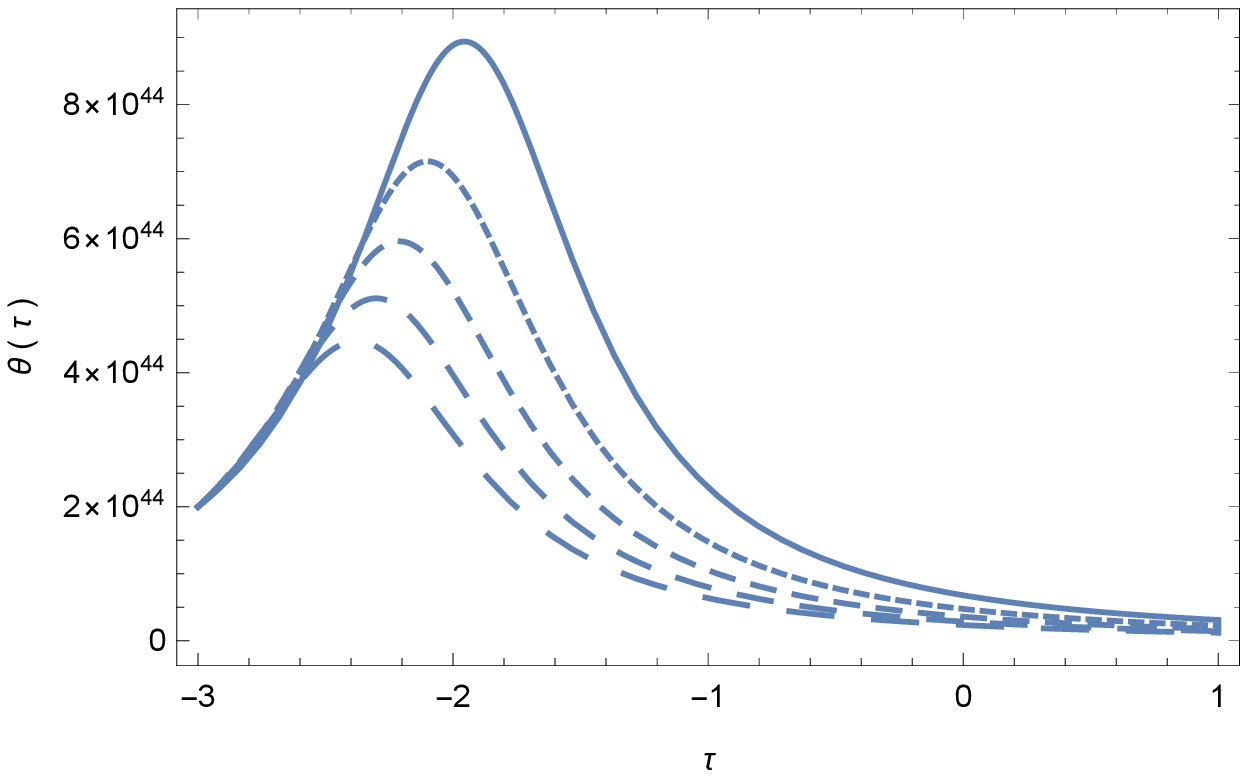}
\includegraphics[width=85 mm]{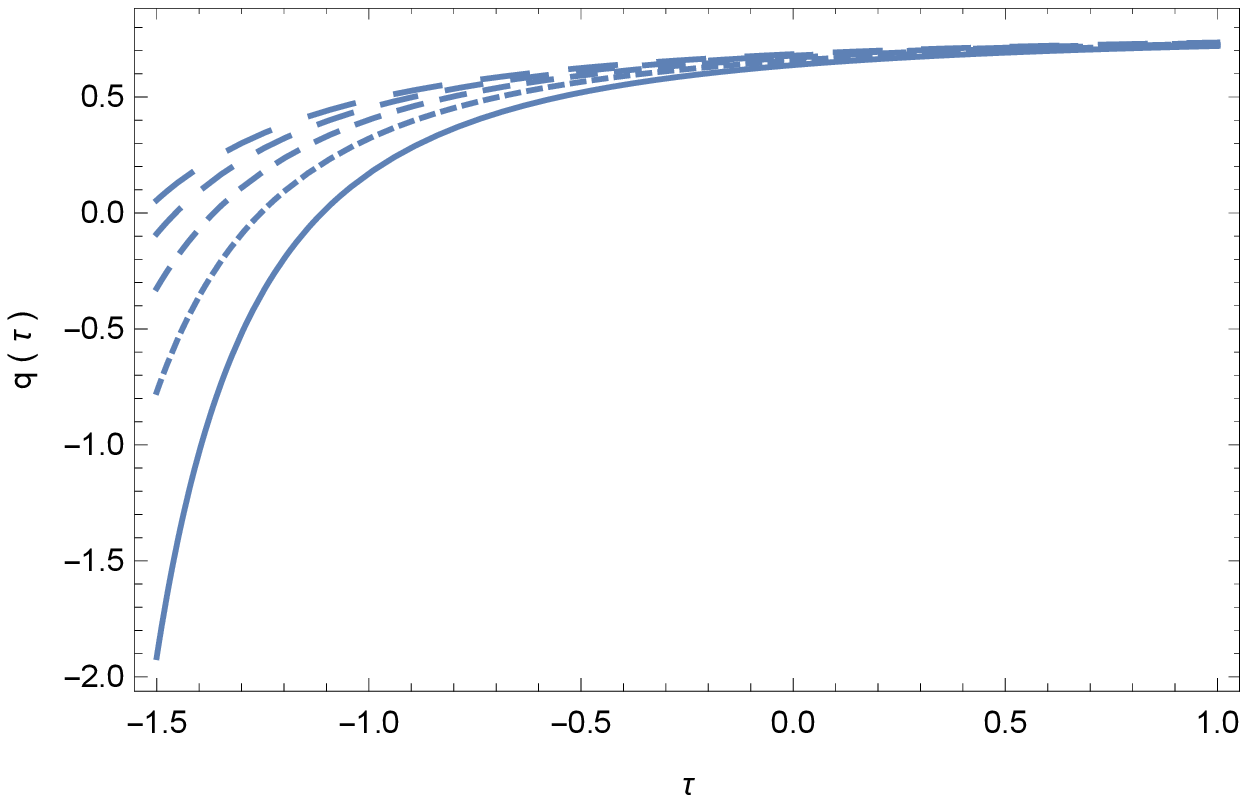}
\caption{Time evolution of the dimensionless temperature $\theta $ (left figure) and of the deceleration parameter $q$ (right figure)in a Loop Quantum Cosmological Universe filled with a deformed photon gas satisfying the dispersion relation $E=kc\sqrt{1-2\beta k^{2}} $, for different values of the parameter $A_3$: $A_3=4\times 10^{-44}$ (solid curve), $A_3=5\times 10^{-44}$ (dotted curve), $A_3=6\times 10^{-44}$ (short dashed curve), $A_3=7\times 10^{-44}$ (dashed curve), and $A_3=8\times 10^{-44}$ (long dashed curve), respectively. The initial conditions used to numerically integrate Eqs.~(\ref{mod21}) and (\ref{mod22}) are $\theta \left(\tau_0\right)=2\times 10^{44}$, and with $h_0$ obtained from Eq.~(\ref{87}). }\label{fig7}
\end{figure*}

The time evolution of $a $, $h$, $\theta$ and of the deceleration parameter $q$ are investigated numerically, for different values of $A_3$. The results of the numerical analysis are presented in Figs.~\ref{fig6} and \ref{fig7}, respectively.

The cosmological evolution is qualitatively similar to the previously considered LQC models. The Universe begins its evolution in a contracting phase, with the scale factor, depicted in the left panel of Fig.~\ref{fig6},  reaching a finite minimum value at some time $\tau _{min}$, after which the Universe bounces, and enters in an expansionary phase, in which the evolution of $a$ is strongly dependent on the numerical values of the parameter $A_3$. The Hubble function, shown in the right panel of Fig.~\ref{fig6}, decreases during the contracting phase, and after reaching a local minimum it increases to a maximum value, followed by a monotonically decreasing phase. The temperature of the Universe, plotted in the left panel of Fig.~\ref{fig7}, increases during the contracting phase, with the Universe reaching its maximum (Planck) temperature at the bounce. After that the temperature as well as the radiation energy density become some decreasing functions of the cosmic time. The time evolution of the deceleration parameter, presented in the right panel of Fig.~\ref{fig7}, shows that in the contracting phase the Universe is strongly accelerating, with the deceleration parameter taking very high (negative values). However, after the bounce, the deceleration parameter increases, and enters in the positive values region, with values of the order of $q\approx 0.5$, indicating a decelerating behavior. In the large time limit the variation of $q$ is independent of the numerical values of the parameter $A_3$.

\subsection{LQC implications of the dispersion relation $E=kc(1-\alpha k)$}

As the third example of a cosmological model in LQC with a modified photon gas
we consider the case in which the deformed dispersion relation for photons
is given by
\begin{equation}\label{disp3}
E=kc(1-\alpha k)  ,
\end{equation}%
where $\alpha $ is a constant. This dispersion relation is particular case
of the more general relation \cite{nc2, ali}
\begin{equation}
E^{2}=k^{2}c^{2}(1-\alpha k)^{2}+m^{2}c^{4}.
\end{equation}%

\subsubsection{Energy density and pressure of the radiation fluid}

For the case of the massless particles, with $m=0$, the energy density of
the photon gas is given by
\begin{equation}
u=\frac{1}{\pi ^{2}\hbar ^{3}}\int_{0}^{\frac{1}{2\alpha }}\frac{k^{2}E}{
(1-2\alpha k+4\alpha ^{2}k^{2})(e^{E/k_{B}T}-1)}dk.
\end{equation}

There are two things one should be mentioned about this integral expression: first, due to the modification of the invariant phase space, the classical degree of degeneracy of each phase unit $\frac{d^{3}\vec{x}d^{3}\vec{k}}{\hbar^{3}}$ is modified to
\be
 \frac{d^{3}\vec{x}d^{3}\vec{k}}{1-2\alpha k+4\alpha ^{2}k^{2}}.
  \ee
 The derivation of this relation is presented in detail in  Appendix ~\ref{app3}. Secondly, the upper limit of integration is taken to be $1/2\alpha $. When $k=1/2\alpha $, $E=E_{max}$; when $k=1/\alpha $, $E=0$. The correct upper bound for the photon momentum is thus $k=1/2\alpha $. If the upper bound of the momentum is taken as $k=1/\alpha $, the pressure will again become negative, which is not physical.

By introducing a new variable $x=kc/k_{B}T$, and defining
\be
\theta=\frac{\alpha k_{B}T}{c},
\ee
the energy density of the photon gas can be written as
\begin{equation}
\begin{aligned}
&u=\frac{k_{B}^{4}T^{4}}{\pi ^{2}\hbar ^{3} c^{3}}\int_{0}^{\frac{1}{2\theta}}\frac{x^{3}(1-\theta x)}{
(1-2\theta x+4\theta ^{2}x^{2})\left[e^{x(1-\theta x)}-1\right]}dx\\&=\frac{k_{B}^{4}T^{4}}{\pi ^{2}\hbar ^{3} c^{3}}M(\theta)=\frac{c}{\alpha^4 \pi^2\hbar^3}\theta^4 M(\theta).
\end{aligned}
\end{equation}
where we have denoted
\begin{equation}
M(\theta)=\int_{0}^{\frac{1}{2\theta}}\frac{x^{3}(1-\theta x)}{
(1-2\theta x+4\theta ^{2}x^{2})\left[e^{x(1-\theta x)}-1\right]}dx.
\end{equation}

The pressure of the deformed radiation gas is given by
\begin{equation}
p=\frac{1}{3\pi ^{2}\hbar ^{3}}\int_{0}^{1/2\alpha }dk \frac{k^{3}}{(1-2\alpha k+4\alpha ^{2}k^{2})(e^{\frac{E}{k_{B}T}}-1)}\frac{dE}{dk},
\end{equation}%
or, equivalently,
\begin{equation}
\begin{aligned}
&p=\frac{k_{B}^{4}T^{4}}{\pi ^{2}\hbar ^{3} c^{3}}\int_{0}^{\frac{1}{2\theta}}\frac{x^{3}(1-2\theta x)}{3(1-2\theta x+4\theta^{2}x^{2})(e^{x(1-\theta x)}-1)}dx\\&=\frac{c}{\pi^2\hbar^3\alpha^4}\theta^4 N(\theta),
\end{aligned}
\end{equation}
where we have introduced the function $N\theta)$, defined as
\begin{equation}
\begin{aligned}
N(\theta)=\int_{0}^{\frac{1}{2\theta}}\frac{x^{3}(1-2\theta x)}{3(1-2\theta x+4\theta^{2}x^{2})(e^{x(1-\theta x)}-1)}dx.
\end{aligned}
\end{equation}

\begin{figure*}[htp]
\centering
\includegraphics[width=85 mm]{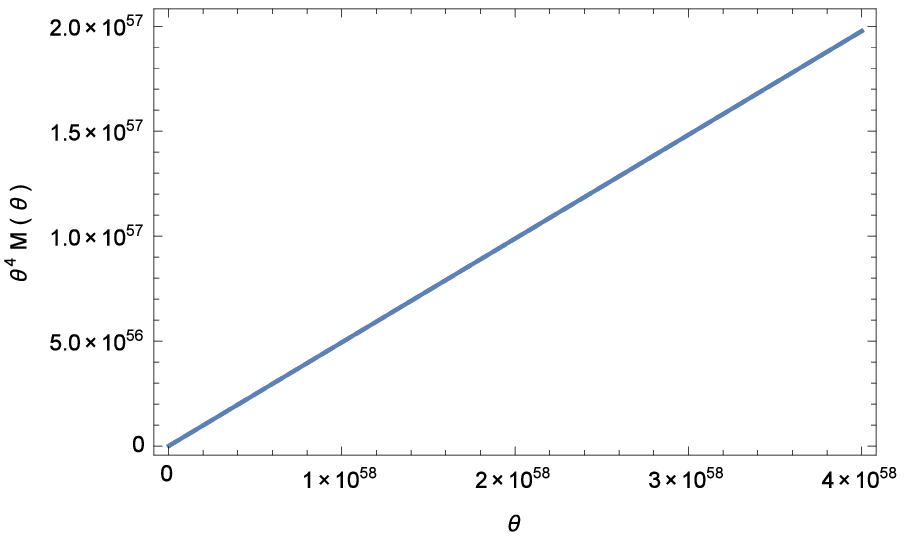}
\includegraphics[width=85 mm]{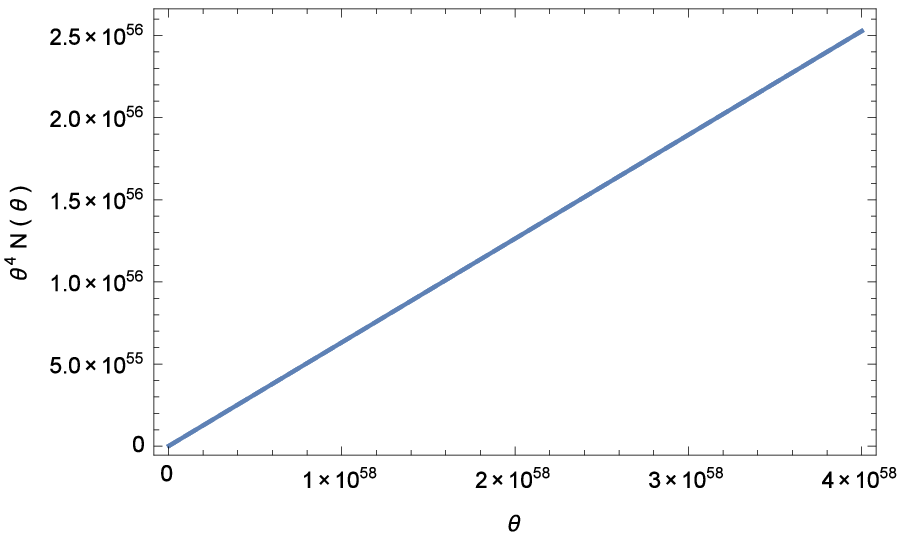}
\caption{Variations of the functions $\theta^4 M(\theta)$ (dotted curve) and $0.04943 \theta$ (solid curve) (left figure) and of the functions $\theta^4 N(\theta)$(dotted curve) and  $0.006317\theta$ (solid curve) (right figure).}\label{fig8}
\end{figure*}

The variations of the functions $\theta^4 M(\theta)$ and $\theta^4 N(\theta)$ are represented in Fig.~\ref{fig8}.

The numerical results can be fitted with a high precision by two linear functions, proportional to $\theta$. Hence we can represent the energy density and the pressure of the photon gas with the modified dispersion relation $E=kc(1-\alpha k)$ as
\be
u=\frac{0.04943 }{\alpha^4 \pi^2\hbar^3}\alpha k_BT,\;
p=\frac{0.006317}{\pi^2\hbar^3\alpha^4}\alpha k_BT.
\ee

\subsubsection{Evolution equations for the radiation filled Universe with photon dispersion relation $E=kc(1-\alpha k)$}

\paragraph{The dimensionless equations of state.} We investigate now the cosmological evolution of Loop
Quantum Cosmology models with deformed radiation described by the modified dispersion relation $E=kc(1-\alpha k)$. In this case the energy density and the pressure
of the radiation can be expressed in the dimensionless form
\begin{equation}
r=0.04943 A_{4} \theta,\;
P=0.006317 A_{4} \theta
\end{equation}%
where we have denoted
\begin{equation}
A_{4}=\frac{1}{\pi ^{2}\hbar ^{3}\alpha^{4} c\rho _{max}}.
\end{equation}%

\paragraph{The generalized Friedmann equations.} Then Eq.~(\ref{c5}) can be written as
\begin{equation}\label{101}
\frac{dh}{d\tau }=-0.0836205 A_4\theta \left(1-0.09885 A_{4} \theta\right).
\end{equation}%

The energy conservation equation (\ref{c6}) takes the form
\begin{equation}\label{102}
\frac{d\theta }{d\tau }+3.3833\;h\theta =0.
\end{equation}

The expression of the deceleration parameter $q$ can be obtained as
\begin{equation}
q=\frac{1.6917\left(1-0.09885 A_{4} \theta\right)}{ 1-0.04943 A_{4} \theta}-1.
\end{equation}

To solve the cosmological evolution of the radiation fluid in the case $E=kc(1-\alpha k)$ numerically, we need to obtain the value of $A_{4}$.
Substituting the expression of $\rho_{max}$ as given in (\ref{c15}) back into $A_{4}$, the expression of $A_{3}$ can be obtained as
\begin{equation}
A_{4}=\frac{8G^{2}\gamma^{2} }{3\pi \hbar^{2} c^{6}\alpha^{4} },
\end{equation}
which is a dimensionless quantity. In the expression of $A_{4}$, $\alpha =\alpha_{0} /M_{p}c$  \cite{nc2}.
By adopting for $\gamma $ and $\alpha $ the values  $\gamma =0.2375$ and $\alpha_{0} =10^{14}$, respectively,  with the upper bound of $\alpha_{0}$ obtained from $^{87}{\rm Rb}$ cold-atom-recoil experiments \cite{87Rb}, the numerical value of $A_4$ can be estimated as
\begin{equation}
A_4\approx 10^{-58}.
\end{equation}

For the dimensionless variable $\theta $ we obtain
\be
\theta =\frac{\alpha _0}{M_{Pl}c^2}E_{Pl}\frac{T}{T_{Pl}}=\alpha _0\frac{T}{T_{Pl}}.
\ee

\subsubsection{The cosmological dynamics}

\begin{figure*}[htp]
\centering
\includegraphics[width=85 mm]{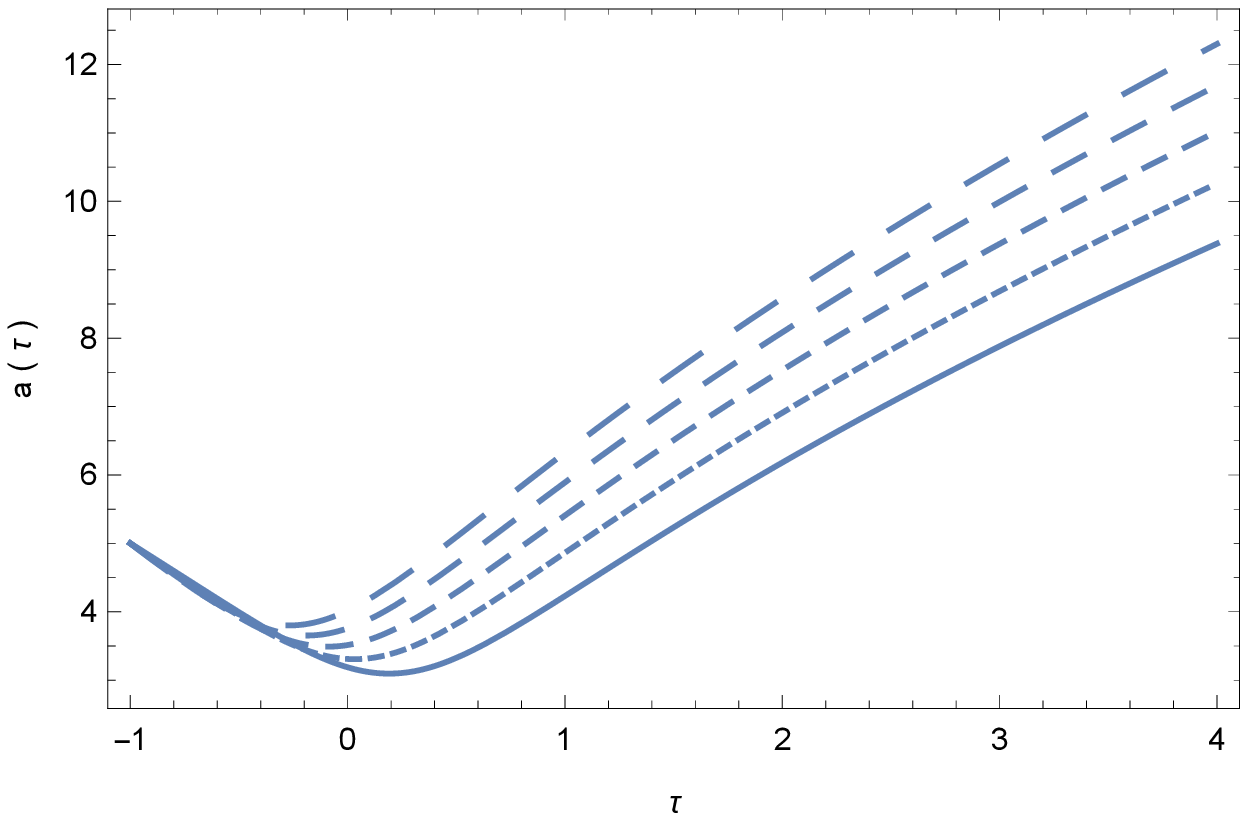}
\includegraphics[width=85 mm]{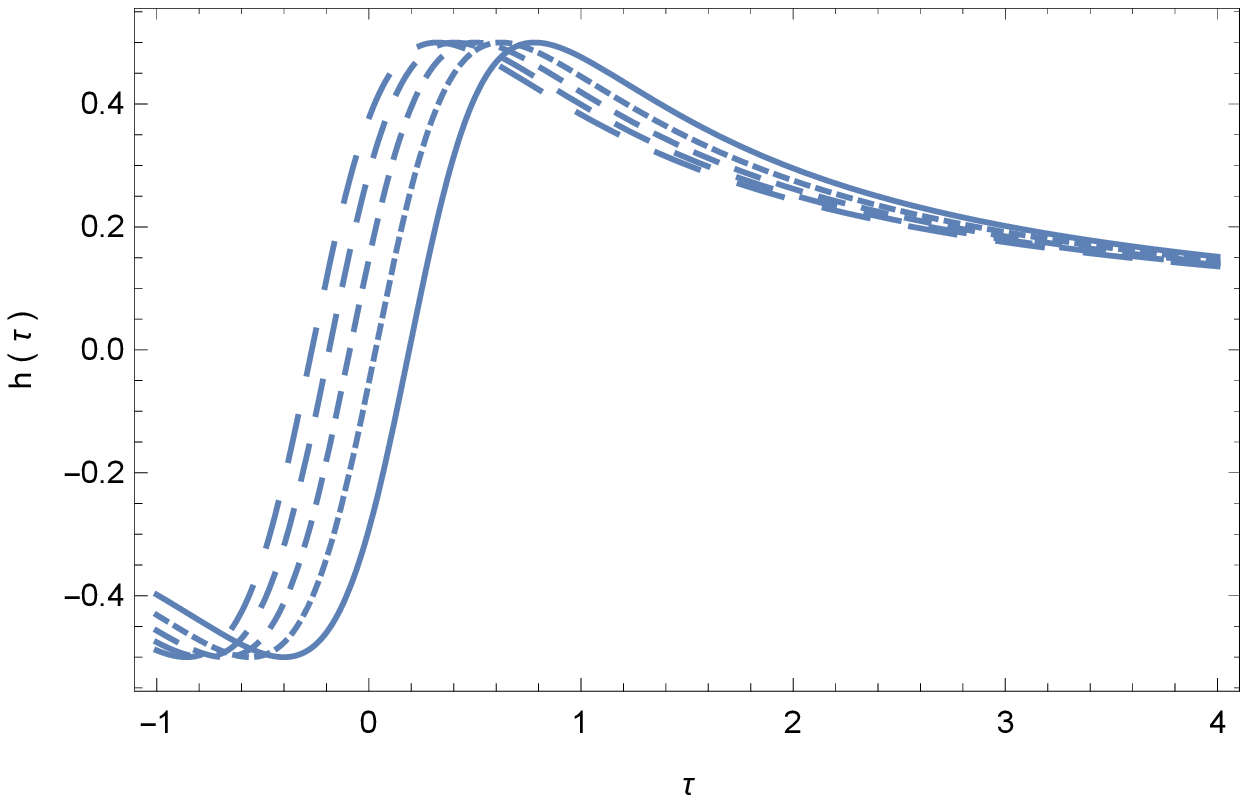}
\caption{Time evolution of the scale factor $a$ (left figure) and of the dimensionless Hubble function $h$ (right figure) in a Loop Quantum Cosmological Universe filled with a deformed photon gas satisfying the dispersion relation  $E=kc(1-\alpha k)$, for different values of the parameter $A_4$: $A_4=4\times 10^{-58}$ (solid curve), $A_4=5\times 10^{-58}$ (dotted curve), $A_4=6\times 10^{-58}$ (short dashed curve), $A_4=7\times 10^{-58}$ (dashed curve), and $A_4=8\times 10^{-58}$ (long dashed curve), respectively. The initial conditions used to numerically integrate Eqs.~(\ref{mod21}) and (\ref{mod22}) are $\theta \left(\tau_0\right)=\theta _0=2\times 10^{58}$, and $h_0=-\sqrt{0.04943 A_{4} \theta _0\left(1-0.04943 A_{4} \theta_0\right)}$, respectively. }\label{fig9}
\end{figure*}

\begin{figure*}[htp]
\centering
\includegraphics[width=85 mm]{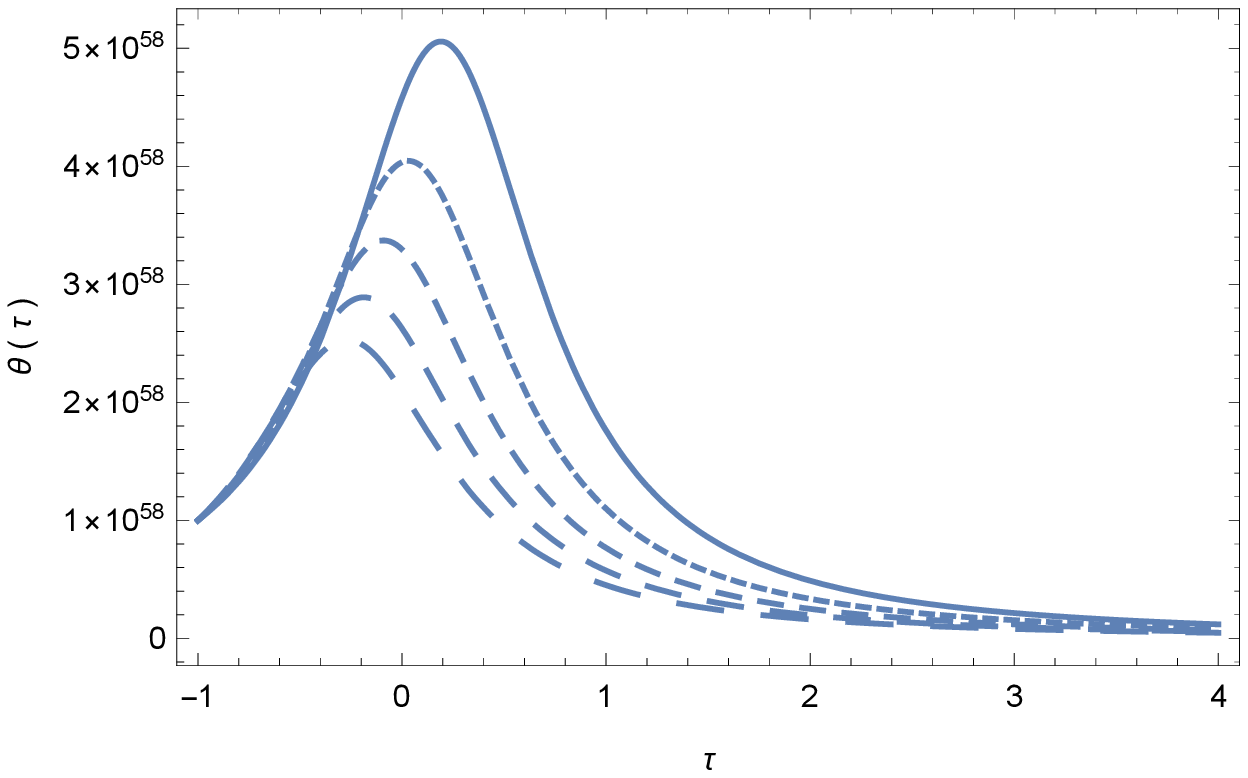}
\includegraphics[width=85 mm]{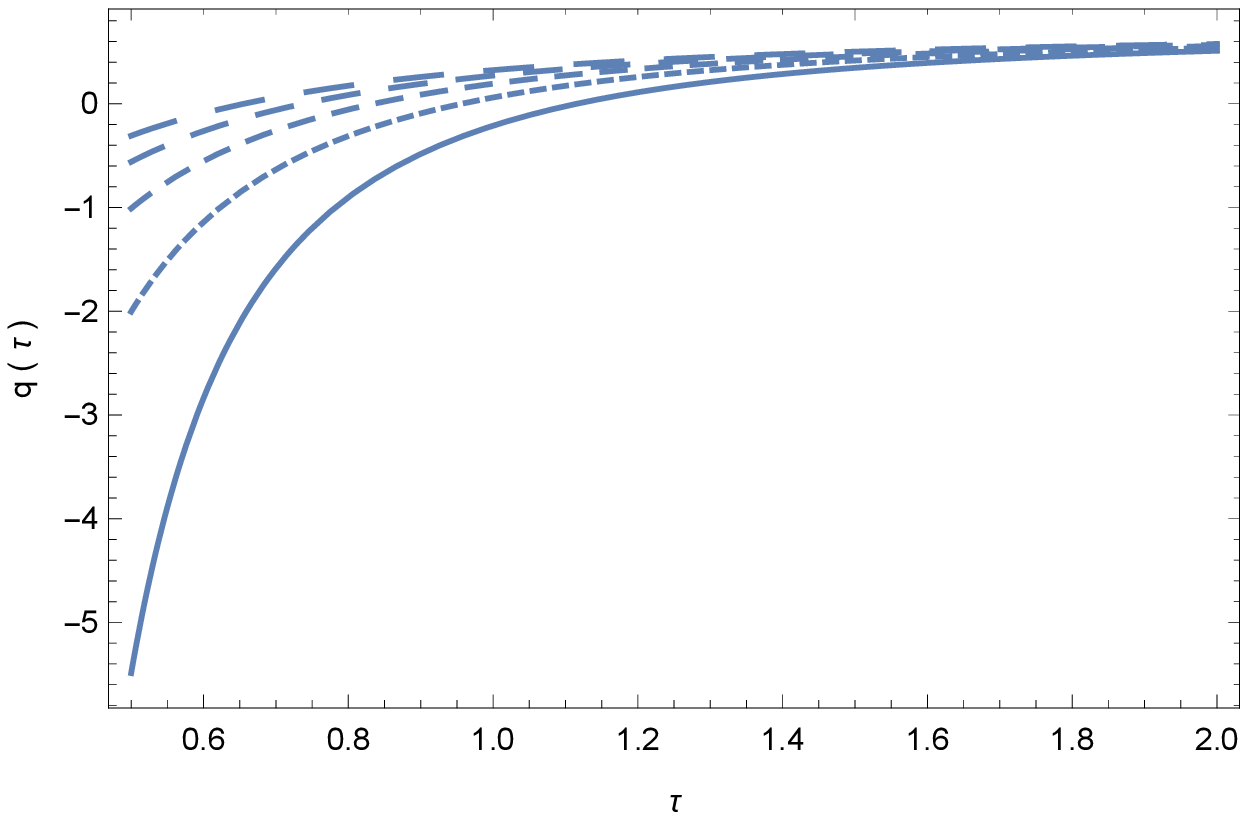}
\caption{Time evolution of the dimensionless temperature $\theta $ (left figure) and of the deceleration parameter $q$ (right figure) in a Loop Quantum Cosmological Universe filled with a deformed photon gas satisfying the dispersion relation  $E=kc(1-\alpha k)$, for different values of the parameter $A_4$: $A_4=4\times 10^{-58}$ (solid curve), $A_4=5\times 10^{-58}$ (dotted curve), $A_4=6\times 10^{-58}$ (short dashed curve), $A_4=7\times 10^{-58}$ (dashed curve), and $A_4=8\times 10^{-58}$ (long dashed curve), respectively. The initial conditions used to numerically integrate Eqs.~(\ref{mod21}) and (\ref{mod22}) are $\theta \left(\tau_0\right)=\theta _0=2\times 10^{58}$, and $h_0=\sqrt{0.04943 A_{4} \theta_0\left(1-0.04943 A_{4} \theta_0\right)}$, respectively. }\label{fig10}
\end{figure*}

The time evolutions of $a $, $h$, $\theta$ and of the deceleration parameter $q$,  obtained by numerically integrating the generalized Friedmann equations (\ref{101}) and (\ref{102}) for different values of the parameter $A_4$ are presented in Figs.~\ref{fig9} and \ref{fig10}, respectively.

The cosmological evolution is qualitatively very similar to the previously considered LQC models, so we will not discuss it in detail. There is a typical bounce at a finite time, with the scale factor and the Hubble function reaching a local minimum, followed by a transition to an expanding state. However, despite the qualitative similarities, there are significant quantitative differences between the different models we have considered. In the present case, for a Universe filled with a radiation fluid satisfying the dispersion relation $E=kc(1-\alpha k)$, the Universe ends in a marginally inflating phase, with $q\approx 0$. The post-bounce behavior of the deceleration parameter is practically independent on the numerical values of the model parameter $A_4$.  The photon gas reaches at the bounce the Planck temperature, which rapidly decreases in the post-bounce phase.

\section{Discussions and final remarks}\label{sect5}

In the present work we have investigated the very early stages of the cosmological evolution of the Universe. Since in this phase the energy scale of the matter reaches the Planck scale, the quantum effects of the gravity must be taken into consideration. Since a complete theory of quantum gravity is still missing, to describe the cosmological evolution at the Planck sale we have adopted the effective LQC approach that results in a set of modified Friedmann equations. Another class of quantum effects on the space time geometry,the  non-commutative nature of the space-time at the Planck scale are also taken into consideration by using, for obtaining the thermodynamic parameters of the photon gas,  the statistical mechanics of the deformed dispersion relations. Non-commutative space-time changes the statistical mechanics by modifying the photon dispersion relation, by restricting the range of the momentum by introducing an upper bound, and by modifying the invariant phase space, respectively.

From a cosmological perspective we have adopted the fundamental assumption that the early Universe consisted of a photon gas only. This physical model has been much less investigated in Loop Quantum Cosmology as compared to the alternative model in which  the early composition of the Universe is modeled by a scalar field only \cite{Asthekar3, Asthekar4, Asthekar5}. However, it is well known that at some moment the Universe was dominated by radiation \cite{Fix}, and we can see the remnants of this radiation in the form of the CMB. In standard cosmology all the matter content of the Universe was produced during the post-inflationary reheating phase by the decay of the inflaton field \cite{reh}. However, the idea of a Universe that may have been dominated from the very early beginning by quantum deformed radiation may represent an attractive alternative to the standard cosmological scenarios implying the presence of a primordial scalar field.

The first drastic effect of the noncommutativity of the space-time on the photon gas at the Planck scale is the modification of the equation of state, and especially of the temperature dependence of the fundamental thermodynamic parameters, the energy density and pressure. In standard quantum physics the radiation gas with a black body spectrum is described by the equation of state $p=u/3$, and with $u\propto T^4$.  In the first model we have discussed,  the dispersion relation is given by $E=kc(1+\lambda E)$, and we have discussed its thermodynamic properties in the limits $\lambda k_{B}T>>1$ and $\lambda k_{B}T<<1$. In the high temperature limit the quantum properties of the space time modify the equation of state of the photons to $u\propto T$, and $\propto T\left(\ln T+A\right)$, where $A$ is a constant. This equation of state is valid for the photon gas at the Planck temperature. In the opposite limit of "small" temperatures, $T\approx 0.1 T_{Pl}$, corrections to the standard photon equation of state do also appear, so that $u\propto \left(\pi ^2/15-AT\right)T^4$, and $p\propto u\left(1/3+BT\right)$, where $A$ and $B$ are constants. The second model we have considered is described by the dispersion relation $E=kc\sqrt{1-2\beta k^{2}}$, with the upper bound of momentum given by $1/2\sqrt{\beta }$, while the invariant phase space is given by $\frac{d^{3}xd^{3}k}{(1+\beta k^{2})^{3}}$. In this case near the Planck temperature the thermodynamic parameters of the photon gas take the simple form $u\propto T$, $p\propto T$, and $p\propto u$, with the proportionality constant in the equation of state different from $1/3$. The last model is described by the dispersion relation $E=kc(1-\alpha k)$, with the upper bound of the momentum given by $1/2\alpha $, and the invariant phase space obtained as $\frac{d^{3}xd^{3}k}{1-2\alpha k+4\alpha^{2} k^{2}}$. The behavior of the thermodynamic parameters is similar to the second model, with both radiation energy density and pressure proportional to the temperature, but with different proportionality coefficients. Even that the equations of state of the radiation fluid can be represented in a closed analytical form in all three cases, we have followed an alternative numerical approach, which is based on the high precision fit of the exact numerical results.

For all the above described equations of state we have studied numerically the modified Friedmann equations of the Loop Quantum Cosmology. In all cases the behavior of the Universe shows qualitatively similar features. We began the investigation of the evolution of the Universe by fixing the initial conditions, which implies giving the values of the initial temperature and Hubble function of the Universe. These initial values fully determine the dynamics of the Universe. By assuming a negative value for $H\left(\tau _0\right)$ (contracting initial state), the bouncing evolution of the radiation gas naturally follows from the LQC modified Friedmann equations. In all the considered models the presence of a singular (geometric or physical) state is completely avoided, and the contracting phase is always followed by an expansionary one. While the scale factor has only one local minimum (the bounce), the Hubble function has both a local minimum and a local maximum. The moment of the maximum of the Hubble parameter occurs later as compared to the moment of the bounce of the scale factor. The energy density of the radiation, which also gives the temperature of the Universe, reaches its maximum value at the bounce. Generally, the behavior of all geometrical and physical quantities show a strong dependence on the numerical values of the basic dimensionless model parameter that combines the noncommutativity and the LQC parameters.

Even that qualitatively similar, the analyzed LQC cosmological models show important quantitative differences. These differences can be seen very clearly by analyzing the behavior of the deceleration parameter. As a common feature of all models, the contracting phase is a super-accelerating one, with $q$ having very high negative values, of the order of $q\approx -10^8$. However, the Universe decelerates quickly, and in the post-bounce phase it is still in an accelerating, de Sitter type phase, with $q$ of the order of $q\approx -1$. The deceleration process continues, and in the large time limit the Universe enters in the positive deceleration parameters phase. This general picture is strongly dependent on the type of the noncommutativity, as well as of the model parameters. In the present analysis, far from the bounce, the deceleration parameter can take values in the range $q\in[0,3]$, and by slightly modifying the model parameters other ranges are also possible. Hence, once the behavior and the numerical values of the deceleration parameter in the very early Universe are known, the comparison of the observational data and the theoretical predictions could lead  to the testing of quantum gravity and noncommutative space-time theories at the Planck scale.  In the framework of a simple physical picture we may interpret the highly accelerating phase before the bounce as kind of pre-inflationary era that is necessary to "push" the Universe in the post-bounce inflationary phase. In the present model at the bounce and immediately after the Universe is still accelerating at a de Sitter rate, and this acceleration is the consequence of the high acceleration acquired in the previous phase, which can be described as a kind of "inertial" behavior when experiencing the transition from one phase to another.

One of the important tests of cosmological models is the study of their linear perturbations. The quantitative predictions of primordial curvature perturbations in LQC depend on the adopted cosmological scenario, and on the initial conditions. Generally,  the evolution of the background can be divided into three different phases, bouncing, transition and slow-roll inflation. In the present analysis we have imposed some initial conditions for the cosmological evolution in the contracting phase, before the bounce. With this choice all the modes are inside the Hubble horizon and the Bunch-Davies  vacuum for perturbations can be imposed. There are some other possibilities for fixing the initial conditions, like, for example,  at the bounce, corresponding to the appearance of the Big Bang "singularity". However,
the Bunch-Davies vacuum cannot be imposed here, as some modes are outside of the Hubble horizon. Another possibility is to adopt as the initial moment the so-called silent point, where $\rho = \rho_{max}/2$.  At this point the perturbations in the algebraic approach change signature (the
hyperbolic equation becomes elliptic for $\rho < \rho_{max}/2)$ \cite{AnWa}. Moreover, it was found that in the dressed  metric approach of \cite{11}  imposing the initial conditions in the contracting phase and at the bounce lead to the same results \cite{AnWa}.

The evolution of the matter density and curvature perturbations in inflation and in the matter bounce scenarios are quite different \cite{testing}.
 Loop Quantum Cosmology models naturally induce an inflationary phase, in which the primeval seed perturbations are created \cite{primordial}. The LQC evolution generates excited states at the onset of the slow-roll inflation \cite{gaussianity}. The power spectrum of density perturbation generated during the induced inflationary phase is broadly scale-invariant \cite{primordial}. There are two scales that determine the form of the LQC-corrected power spectrum. The first one is associated with the spacetime curvature at the bounce ($k_{I}$), while the second one is associated with the spacetime curvature at the onset of inflation ($k_{LQC}$).  The amplitude of the power spectrum is amplified with respect to the standard predictions of slow-roll inflation for modes $k_{I}<k<k_{LQC}$ \cite{detailed}. The initial value of the inflaton field (which is a massive scalar field) $\psi_{B}$, and the mass of the massive scalar field $m_{\phi}$ control whether the modes affected by the pre-inflationary dynamics of LQC fall within the window of modes that are observable today. The region that shows non-negligible LQC modifications is a subset which satisfied the condition $\psi_{B}\approx 3\times 10^{-6}/m_{\phi} $ \cite{detailed}. LQC corrections to the power spectra tend to make the tensor spectral index $n_{t}$ more negative, produce a positive running $\alpha_{s}$ of the scalar spectral index, reduce the tensor-to-scalar ratio $r$, and modify the inflationary consistency relation to $r/n_{t}=-8$ \cite{detailed}. The particular choice of the initial data for the quantum scalar and tensor perturbations has very little impact on the above conclusions, at least for the reasonable choices of initial vacuum state that is considered above \cite{detailed}.

If we only consider the suitable initial vacuum state that satisfy the requirement that the initial conditions for each mode minimize the time variation of the Mukhanov-Sasaki variable from the bounce to the beginning of the inflation, LQC can generate a power spectrum of the curvature perturbations in which large oscillations with the averaged enhanced power are not present. Instead, in LQC one can  obtain for that range of scales a behaviour compatible with the one obtained from the slow-roll inflation formula. Moreover, LQC also predicts a strong power suppression for large scales, a result that provides a better fitting of the current observations \cite{Daniel,testing}. The same power suppression effect is predicted to occur in the power spectrum of tensor modes as well \cite{testing}.

The non-Gaussianity of the primordial curvature perturbations in Loop Quantum Cosmology have been also studied, and it was found that the corrections due to quantum effects are of the same order of magnitude as the inflationary slow-roll parameters in the observable scales. Thus they are well within current observational constraints \cite{gaussianity}. The non-Gaussianity in the squeezed limit can be enhanced at superhorizon scales. These effects may yield a large statistical anisotropy on the power spectrum \cite{gaussianity}.

In the matter bounce scenario, the fluctuation of the curvature and the tensor perturbations become scale-invariant in a contracting FLRW space-time, where the matter content has vanishing pressure \cite{testing}. Then, if a bounce can be generated to provide a non-singular transition from contraction to expansion, these scale-invariant perturbations provide suitable initial conditions for the expanding Universe that can explain the scale-invariance observed in the CMB, under the assumption that the bounce does not modify the power spectrum \cite{testing}. LQC can generate the bounce that is required for this scenario to be viable \cite{testing}. During the LQC predicted bounce, the tensor-to-scalar ratio will be suppressed. The precise numerical factor of the tensor-to-scalar ratio suppression depends on the dominant matter fields during the bounce. The possible observational dependence of the suppression factor may provide important information about the physics of the bounce \cite{testing}.

In the present approach of a radiation filled Loop Quantum Cosmological Universe, one of the theoretical possibilities is  that the primordial curvature perturbations are generated during the induced inflation period. Hence we may assume that the generation of the primordial curvature perturbation takes place in the post-bounce phase. However, the presence of the matter bounce makes also possible the generation of the primordial curvature perturbations in  the early contracting phase. During the contraction towards the bounce, the vacuum fluctuations and the curvature perturbations become scale-invariant. Then the existence of the non-singular bounce predicted by LQC leads to the transition from contraction to expansion, a process that does not modify the scale-invariance of the power spectrum. The scale-invariant power spectrum, maintained after the bounce, provides a suitable initial condition for the expanding Universe, which matches our observations based on the Cosmic Microwave Background Radiation.

If the primordial curvature perturbation are explained by the post-bounce inflationary scenario, the accelerating state before the bounce can be considered as a pre-inflationary era, whose main role is to generate excited states at the onset of inflation. Hence, in this scenario, the bouncing state corresponds to the starting point of the pre-inflationary phase. The boundary of the pre-inflation era is characterized by the quantities $k_{LQC}$ and $k_{L}$, which also describe the power spectrum of density perturbations generated during the inflation. If the generation of primordial curvature perturbation is explained during the contracting phase, the accelerating era before the bounce can be considered as the period when the vacuum fluctuations in curvature and tensor perturbations become scale-invariant. In this scenario, the bouncing point, representing the non-singular transition point from the contracting to the expanding era maintains the scale-invariance of the power spectra of the density perturbations.
On the other hand, since the generation of the matter density/curvature perturbations during the inflationary era and during the  matter pre-bounce phase predict different power spectra, and different tensor-to-scalar ratios, it is possible to test the viability of these scenarios via observations of CMB, as described in \cite{testing}.

In the present paper we have introduced a theoretical model that combines two major approaches that may lead to a better description of the quantum properties of the gravitational force, Loop Quantum Cosmology and noncommutative space-time theories, respectively. These results may open the possibility of a better understanding of the theoretical phenomena at the Planck scale, and of the in depth comparison of the theoretical predictions with the observational data.

\section*{Acknowledgments}

We would like to thank to the two anonymous reviewers for comments and suggestions that helped us to significantly improve our manuscript. T.H. would like to thank the Yat-Sen School of the Sun Yat-Sen University in
Guangzhou, P. R. China, for the kind hospitality offered during the
preparation of this work. S.-D. Liang acknowledges the support
of the Natural Science Foundation of Guangdong Province (Grant no.
2016A030313313).


\appendix

\section{Energy density and pressure of radiation fluid satisfying the dispersion relation $E=kc\sqrt{1-2\beta k^{2}}$ }\label{app1}

In order to obtain an exact analytic representation  of the integral in Eq.~(\ref{75}) we introduce first the transformation $x=\frac{1}{\sqrt{2\tilde{T}}}\sin\theta $, which gives $dx=\frac{1}{\sqrt{2\tilde{T}}}\cos\theta d\theta $.
Then
\bea
\hspace{-0.6cm}&&I\left(\tilde{T}\right)=\nonumber\\
\hspace{-0.6cm}&&\frac{1}{\left(2\tilde{T}\right)^{2}}\sum_{n=0}^{\infty}\int_{0}^{\pi/4}\frac{\sin^{3}\theta \cos^{2}\theta}{
\left[1+(1/2)\sin^{2}\theta \right]^{3}}e^{-\frac{(n+1)}{\sqrt{2\tilde{T}}}\sin\theta \cos\theta }d\theta ,
\eea
\begin{equation}
e^{-\frac{(n+1)}{\sqrt{2\tilde{T}}}\sin\theta \cos\theta }=\sum_{m=0}^{\infty}(-1)^{m}\frac{(n+1)^{m}}{\left(2\tilde{T}\right)^{m/2}}\sin^{m}\theta \cos^{m}\theta ,
\end{equation}
\begin{equation}
I\left(\tilde{T}\right)=\sum_{n,m=0}^{\infty}(-1)^{m}\frac{(n+1)^{m}}{\left(2\tilde{T}\right)^{m/2+2}}\int_{0}^{\pi/4}\frac{\sin^{3+m}\theta \cos^{2+m}\theta}{\left[1+(1/2)\sin^{2}\theta \right]^{3}}d\theta .
\end{equation}
Therefore the energy density of the deformed radiation fluid  is obtained as
\begin{equation}
u=\frac{(k_{B}T)^{4}}{\pi ^{2}\hbar ^{3} c^{3}}\sum_{n,m=0}^{\infty}\frac{(-1)^{m}(n+1)^{m}}{\left(2\tilde{T}\right)^{m/2+2}}F(m),
\end{equation}
where we have denoted
\be
F(m)=\int_{0}^{\pi/4}\frac{\sin^{3+m}\theta \cos^{2+m}\theta}{\left[1+(1/2)\sin^{2}\theta \right]^{3}}d\theta .
\ee

The pressure of the deformed radiation gas is given by
\begin{equation}
p=\frac{1}{3\pi ^{2}\hbar ^{3}}\int_{0}^{\frac{1}{2\sqrt{\beta }}}dk \frac{k^{3}}{\left(1+\beta k^{2}\right)^{3}\left(e^{\frac{E}{k_{B}T}}-1\right)}\frac{dE}{dk},
\end{equation}%
or, equivalently,
\bea
\hspace{-0.4cm}&&p=\frac{u}{3}-\frac{k_{B}^{4}T^{4}}{\pi ^{2}\hbar ^{3} c^{3}}\frac{2\tilde{T}}{3}\times\nonumber\\
\hspace{-0.4cm}&&\int_{0}^{\frac{1}{2\sqrt{\tilde{T}}}}\frac{x^{5}}{(1+\tilde{T}x^{2})^{3}\sqrt{1-2\tilde{T}x^{2}}(e^{x\sqrt{1-2\tilde{T}x^{2}}}-1)}dx.
\eea

The integral in the expression of pressure can also be written in the form of series expansion, which is
\begin{equation}
p=\frac{u}{3}-\frac{2}{3}\frac{k_{B}^{4}T^{4}}{\pi ^{2}\hbar ^{3} c^{3}}\sum_{n,m=0}^{\infty}\frac{(-1)^{m}(n+1)^{m}}{\left(2\tilde{T}\right)^{m/2+2}}G(m),
\end{equation}
in which
\begin{equation}
G(m)=\frac{1}{2}\int_{0}^{\pi/4}\frac{\sin^{5+m}\theta \cos^{m}\theta }{(1+\frac{\sin^{2}\theta }{2})^{3}}d\theta .
\end{equation}

\section{Calculation of the invariant phase space of the deformed radiation gas with  $E=kc(1-\alpha k)$}\label{app3}

When the dispersion relation of deformed radiation is given by $E=kc(1-\alpha k)$, the modified Heisenberg relation is \cite{nc2}
\begin{equation}
[x_{i},p_{j}]=i\hbar \left[\delta_{ij}-\alpha \left(p\delta_{ij} +p_{i}p_{j}/p\right)+\alpha^{2} \left(p^{2}\delta_{ij} +3p_{i}p_{j}\right)\right].
\end{equation}
The components of the momentum $p_{i}$, $i=1,2,3$ are assumed to be commute with each other \cite{ips}
\begin{equation}
\left[p_{i},p_{j}\right]=0.
\end{equation}
Then the commutators  $\left[x_{i},x_{j}\right]$ can be computed by using the Jacobi identity,
\begin{equation}
\left[p_{k},\left[x_{i},x_{j}\right]\right]+\left[x_{i},\left[x_{j},p_{k}\right]\right]+\left[x_{j},\left[p_{k},x_{i}\right]\right]=0.
\end{equation}
Hence it follows that
\begin{equation}
\left[x_{i},x_{j}\right]=i\hbar \alpha \frac{p_{i}}{p}x_{j}-i\hbar \alpha \frac{p_{j}}{p}x_{i}-3i\hbar \alpha^{2}p_{i}x_{j}+3i\hbar \alpha^{2}p_{j}x_{i}.
\end{equation}
The classical limit of the above commutators is given by
\begin{equation}
\left\{x_{i},p_{j}\right\}=\delta_{ij}-\alpha \left(p\delta_{ij} +p_{i}p_{j}/p\right)+\alpha^{2} \left(p^{2}\delta_{ij} +3p_{i}p_{j}\right),
\end{equation}
\begin{equation}
\left\{p_{i},p_{j}\right\}=0.
\end{equation}
\begin{equation}
\left\{x_{i},x_{j}\right\}=\alpha \frac{p_{i}}{p}x_{j}-\alpha \frac{p_{j}}{p}x_{i}-3\alpha^{2}p_{i}x_{j}+3\alpha^{2}p_{j}x_{i}.
\end{equation}
The time evolution of the coordinate and the momentum are governed by \cite{ips}
\begin{equation}
\frac{dx_{i}}{dt}=\left\{x_{i},H\right\}=\left\{x_{i},p_{i}\right\}\frac{\partial H}{\partial p_{j}}+\left\{x_{i},x_{j}\right\}\frac{\partial H}{\partial x_{j}}
\end{equation}
\begin{equation}
\frac{dp_{i}}{dt}=\left\{p_{i},H\right\}=-\left\{x_{j},p_{i}\right\}\frac{\partial H}{\partial x_{j}}
\end{equation}
To first order approximation in $\delta t$ we obtain \cite{ips}
\begin{equation}
\begin{aligned}
&d^{3}x'd^{3}p'=\left|\frac{\partial (x'_{1},...,x'_{3},p'_{1},...,p'_{3})}{\partial (x_{1},...x_{3},p_{1},...,p_{3})}\right|d^{3}xd^{3}p\\&=\left(1+\frac{\partial \delta x_{i}}{\partial x_{i}}+\frac{\partial \delta p_{i}}{\partial p_{i}}\right)d^{3}xd^{3}p,
\end{aligned}
\end{equation}
and \cite{ips}
\begin{equation}
\begin{aligned}
&\left(\frac{\partial \delta x_{i}}{\partial x_{i}}+\frac{\partial \delta p_{i}}{\partial p_{i}}\right)\frac{1}{\delta t}\\&=\frac{\partial }{\partial x_{i}}\left[\{x_{i},p_{j}\}\frac{\partial H}{\partial p_{j}}+\{x_{i},x_{j}\}\frac{\partial H}{\partial x_{j}}\right]-\frac{\partial }{\partial p_{i}}\left[\{x_{j},p_{i}\}\frac{\partial H}{\partial x_{j}}\right]\\&=\left[\frac{\partial }{\partial x_{i}}\{x_{i},x_{j}\}\right]\frac{\partial H}{\partial x_{j}}-\left[\frac{\partial }{\partial p_{i}}\{x_{j},p_{i}\}\right]\frac{\partial H}{\partial x_{j}},
\end{aligned}
\end{equation}
respectively. Substituting $\left\{x_{i},x_{j}\right\}$ and $\left\{x_{j},p_{i}\right\}$ into the above expression, and noticing that
\begin{equation}
\frac{\partial }{\partial p_{j}}p=\frac{p_{j}}{p},\,\,
\frac{\partial }{\partial p_{i}}\frac{p_{j}p_{i}}{p}=D\frac{p_{j}}{p},
\end{equation}
\begin{equation}
\frac{\partial }{\partial p_{j}}p^{2}=2p_{j},\,\,
\frac{\partial }{\partial p_{i}}p_{j}p_{i}=(D+1)p_{j},
\end{equation}
where $D=3$ is the dimension of the system, it can be shown that
\begin{equation}
\left(\frac{\partial \delta x_{i}}{\partial x_{i}}+\frac{\partial \delta p_{i}}{\partial p_{i}}\right)\frac{1}{\delta t}=\left(\frac{2\alpha }{p}-8\alpha ^{2}\right)p_{j}\frac{\partial H}{\partial x_{j}},
\end{equation}
and
\begin{equation}
d^{3}x'd^{3}p'=d^{3}d^{3}p\left[1+\left(\frac{2\alpha }{p}-8\alpha ^{2}\right)p_{j}\frac{\partial H}{\partial x_{j}}\delta t\right],
\end{equation}
respectively. Hence finding the invariant phase space is equivalent to finding a function $f(p)$, so that
\begin{equation}\label{c21}
f\left(p'\right)=f(p)\left[1+\left(\frac{2\alpha }{p}-8\alpha ^{2}\right)p_{j}\frac{\partial H}{\partial x_{j}}\delta t\right].
\end{equation}
In the first order approximation we have,
\begin{equation}
\begin{aligned}
&f\left(p'\right)=f\left(\sqrt{(p_{i}+\delta p_{i})^{2}}\right)=f\left(\sqrt{p_{i}^{2}+2p_{i}\delta p_{i}}\right)\\&=f(p)+\left.\frac{df(x)}{dx}\right|_{p'}\frac{p_{i}}{p}\delta p_{i}\\&=f(p)+\left(\left.\frac{df(x)}{dx}\right|_{p}+\left.\frac{d^{2}f(x)}{dx^{2}}\right|_{p}\delta p\right)\frac{p_{i}}{p}\delta p_{i}\\&=f(p)+\frac{df(p)}{dp}\frac{p_{i}}{p}\delta p_{i}=f(p)-\frac{df(p)}{dp}\frac{p_{i}}{p}\left\{x_{j},p_{i}\right\}\frac{\partial H}{\partial x_{j}}\delta t.
\end{aligned}
\end{equation}
Substituting $\left\{x_{j},p_{i}\right\}$ into the above expression, we obtain
\begin{equation}\label{c22}
\begin{aligned}
&f\left(p'\right)=f(p)\times \\&\left[1-\frac{1}{f(p)}\left(\frac{df(p)}{dp}\frac{1}{p}-2\alpha \frac{df(p)}{dp}+4\alpha^{2} \frac{df(p)}{dp}p\right)p_{j}\frac{\partial H}{\partial x_{j}}\delta t\right].
\end{aligned}
\end{equation}
Comparing Eqs.~(\ref{c21}) and (\ref{c22}), it follows that $f(p)$ must satisfy the condition
\begin{equation}\label{c23}
\begin{aligned}
\frac{2\alpha }{p}-8\alpha ^{2}=-\frac{1}{f(p)}\left[\frac{df(p)}{dp}\frac{1}{p}-2\alpha \frac{df(p)}{dp}+4\alpha^{2} \frac{df(p)}{dp}p\right].
\end{aligned}
\end{equation}
Eq.~(\ref{c23}) can be simplified as
\begin{equation}
\begin{aligned}
\frac{1}{f(p)}\frac{df(p)}{dp}=\frac{8\alpha^{2} p-2\alpha }{4\alpha^{2} p^{2}-2\alpha p+1},
\end{aligned}
\end{equation}
which can be integrated to give
\begin{equation}
\begin{aligned}
\ln(f(p))=\ln \left[C\left(1-2\alpha p+4\alpha^{2} p^{2}\right)\right],
\end{aligned}
\end{equation}
\\
where $C$ is an arbitrary integration constant.
Thus,
\begin{equation}
\begin{aligned}
f(p)=C\left(1-2\alpha p+4\alpha^{2} p^{2}\right).
\end{aligned}
\end{equation}
In the limit $\alpha =0$, $f(p)=1$, which is the classical Liouville theorem. Hence we must have $C=1$, and thus
\begin{equation}
\begin{aligned}
f(p)=1-2\alpha p+4\alpha^{2} p^{2}.
\end{aligned}
\end{equation}
Therefore for the invariant phase space of the deformed radiation fluid with  $E=kc(1-\alpha k)$ we obtain
\be
 \frac{d^{3}xd^{3}p}{1-2\alpha p+4\alpha^{2} p^{2}}.
 \ee

\end{document}